\documentclass[a4paper,10pt]{article}

\usepackage{jheppub} 

\usepackage[T1]{fontenc} 
\usepackage{bbold}
\usepackage{amsmath}
\usepackage{empheq}
\usepackage{booktabs}
\usepackage{slashed}
\usepackage{url}

\newcommand{\rd}{{\rm d}}

\newcommand{\re}{{\rm e}}

\newcommand{\as}{\alpha_s}
\newcommand{\aw}{\alpha}

\usepackage{xspace}

\newcommand{\rad}{\textsc{RadISH}\xspace}

\newcommand{\matr}{\textsc{MATRIX}\xspace}
\newcommand{\pwg}{\textsc{POWHEG$_{\rm QCD+EW}$}\xspace}
\newcommand{\pwgg}{\textsc{POWHEG}\xspace}
\newcommand{\py}{\textsc{Pythia8}\xspace}
\newcommand{\vin}{\textsc{Vincia}\xspace}
\newcommand{\hw}{\textsc{Herwig7}\xspace}
\newcommand{\sherpa}{\textsc{Sherpa}\xspace}
\newcommand{\horace}{\textsc{Horace}\xspace}
\newcommand{\pho}{\textsc{Photos}\xspace}
\newcommand{\recola}{\textsc{Recola}\xspace}
\newcommand{\eq}[1]{eq.~(\ref{#1})}

\def\({\left(} 
\def\){\right)}

\newcommand{\muR}{\mu_R}
\newcommand{\muF}{\mu_F}
\newcommand{\beq}{\begin{eqnarray}}
\newcommand{\eeq}{\end{eqnarray}}
\newcommand{\nnb}{\nonumber}

\usepackage[dvipsnames]{xcolor}
\usepackage[normalem]{ulem}

\title{\boldmath Resummation of combined QCD-electroweak effects in Drell Yan lepton-pair production}

\author[a]{Luca Buonocore,}
\author[b]{Luca Rottoli,}
\author[c]{and Paolo Torrielli.}

\affiliation[a]{CERN, Theoretical Physics Department, CH-1211 Geneva 23, Switzerland}
\affiliation[b]{Department of Physics, University of Z\"urich, CH-8057 Z\"urich, Switzerland}
\affiliation[c]{Dipartimento di Fisica, Universit\`a di Torino, and INFN, Sezione di Torino,
\\
Via P. Giuria 1, I-10125 Torino, Italy}

\emailAdd{lbuonocor@cern.ch}
\emailAdd{luca.rottoli@physik.uzh.ch}
\emailAdd{torriell@to.infn.it}

\abstract{
We consider neutral- and charged-current Drell Yan lepton-pair production at hadron colliders, and include dominant classes of electroweak and mixed QCD-electroweak corrections to all orders in perturbation theory. The accurate description of these physical effects is vital for a precise determination of fundamental Standard Model parameters, such as the $W$-boson mass and the electroweak mixing angle, as well as for a solid assessment of the associated theoretical uncertainties.
Our state-of-the-art resummation reaches next-to-leading-logarithmic accuracy in both the electroweak and the mixed QCD-electroweak perturbative expansions, including constant terms at first order beyond Born level in both couplings, i.e.~at order  $\aw$ and $\as \aw$. These effects are incorporated on top of QCD predictions at next-to-next-to-next-to-leading-logarithmic accuracy, which include constant terms at third order in the strong coupling. Our results retain, for the first time at this accuracy, full dependence on the kinematics of the final-state leptons, thereby enabling a realistic comparison with experimental analyses at the differential level in presence of fiducial cuts. We present a phenomenological analysis of the impact of electroweak corrections in relevant observables at the LHC. We find visible shape distortions in resummation-dominated kinematical regions with respect to pure-QCD predictions, highlighting the importance of a complete description, not limited to QCD, for precision Drell Yan physics.}

\preprint{
\begin{flushright}
ZU-TH 21/24\\
CERN-TH-2024-045
\end{flushright}
}

\allowdisplaybreaks

\begin{document} 
\maketitle
\flushbottom

\section{Introduction}

The production of lepton pairs through the Drell Yan (DY) mechanism is central to the precision physics programme of hadron colliders such as the Tevatron and the LHC. Owing to its large cross section and its clean experimental signature, with at least one hard charged lepton in the final state, the DY process features prominently in the precise determination of fundamental Standard Model (SM) parameters, such as the $W$-boson mass \cite{Group:2012gb,ATLAS:2017rzl,LHCb:2021bjt,CDF:2022hxs,ATLAS:2023fsi}, the electroweak (EW) mixing angle \cite{CMS:2011utm,ATLAS:2015ihy,LHCb:2015jyu,CDF:2018cnj,CMS:2018ktx,ATLAS:2018gqq,CMS:2024aps}, and the strong coupling~\cite{ATLAS:2023lhg}.
Moreover, it provides stringent constraints on parton distribution functions (PDFs) of the proton, and represents a dominant background for many signals, both within and beyond the SM.

On the theoretical side, it is crucial to provide predictions of the highest accuracy for DY fiducial cross sections and differential distributions. Given the level of accuracy attained nowadays by experimental measurements \cite{ATLAS:2019zci,CMS:2019raw,LHCb:2021huf,CMS:2022ubq,ATLAS:2023lsr,LHCb:2023qav,ATLAS:2024nrd}, not only does this imply the account of QCD radiative corrections at high perturbative orders, but also the inclusion of EW contributions, whether they are pure EW effects or QCD-EW interferences.

At fixed order in QCD perturbation theory, after pioneering
next-to-leading-order (NLO) and next-to-NLO (NNLO) inclusive DY calculations
\cite{Altarelli:1979ub,Hamberg:1990np,Harlander:2002wh}, differential results
became available at NNLO
\cite{Anastasiou:2003yy,Anastasiou:2003ds,Melnikov:2006di,Melnikov:2006kv,Catani:2009sm,Catani:2010en}.
Nowadays, the next-to-NNLO (N$^3$LO) level has been reached both for total rates
\cite{Duhr:2020seh,Duhr:2020sdp,Baglio:2022wzu}, and for differential
distributions
\cite{Chen:2021vtu,Chen:2022cgv,Chen:2022lwc,Chen:2022lpw,Neumann:2022lft,Campbell:2023lcy,Camarda:2021ict}.
The NLO EW corrections have been calculated long ago for charged-current DY (CCDY) in Refs.~\cite{Dittmaier:2001ay,Baur:2004ig,Zykunov:2006yb,Arbuzov:2005dd,CarloniCalame:2006zq}
and for neutral-current DY (NCDY) in
Refs.~\cite{Baur:2001ze,Zykunov:2005tc,CarloniCalame:2007cd,Arbuzov:2007db,Dittmaier:2009cr}.
Efforts to go beyond that level have witnessed a revived interest in recent years,
with the calculation of NNLO mixed QCD-EW effects. First results have been
obtained in the {\it pole} approximation (see Ref.~\cite{Denner:2019vbn} for a
general discussion) with the calculation in
Refs.~\cite{Dittmaier:2014qza,Dittmaier:2015rxo} of the so-called {\it factorisable}
contributions of {\it initial--final} and {\it final--final} type. More recently, the missing
{\it initial--initial} contributions have been considered in
Ref.~\cite{Dittmaier:2024row}. Going beyond the pole approximation, mixed
QCD-QED corrections were obtained in Ref.~\cite{deFlorian:2018wcj,Delto:2019ewv} for the inclusive production of an on-shell $Z$ boson, and in Ref.~\cite{Cieri:2020ikq} for off-shell $Z$ boson production and decay into a pair of neutrinos at the fully differential level. Mixed QCD-EW ${\cal O}(\as\alpha)$ double-real corrections were obtained in \cite{Bonciani:2016wya} for on-shell $Z$ and $W$ production, while the complete ${\cal O}(\as\alpha)$ computation for the production of on-shell $Z$ bosons has been
presented in Refs.~\cite{Bonciani:2019nuy,Bonciani:2020tvf}. For the off-shell
case, there is a computation~\cite{Buonocore:2021rxx} of the mixed QCD-EW
corrections to CCDY, where all contributions are
evaluated exactly except for the finite part of the two-loop amplitude, which
was evaluated in the pole approximation. As for NCDY, thanks to the calculation of its exact two-loop amplitude~\cite{Heller:2020owb,Armadillo:2022bgm}, the complete mixed QCD-EW corrections have been achieved in Refs.~\cite{Bonciani:2021zzf,Buccioni:2022kgy}.

It is well known that predictions at fixed order in perturbation theory are
reliable only for sufficiently inclusive quantities. Whenever an observable is
sensitive to infrared and/or collinear (IRC) radiation, large logarithms arise
in the calculation of its higher-order corrections, featuring as argument the
ratio of a hard to an IRC scale. The presence of such logarithms spoils the
convergence of the perturbative expansion, and claims for a resummation of
logarithmic enhancements to all perturbative orders. In the case of QCD
corrections to the DY process, the resummation of IRC-sensitive observables like
the lepton-pair transverse momentum $p_t^{\ell\ell}$ or the $\phi_\eta^*$
distribution \cite{Banfi:2010cf} has seen a steady evolution, from seminal
papers \cite{Parisi:1979se,Collins:1984kg,Balazs:1995nz,
  Balazs:1997xd,Catani:2000vq} to more recent developments in a variety of
formalisms
\cite{Bozzi:2008bb,Bozzi:2010xn,Becher:2010tm,Banfi:2011dm,GarciaEchevarria:2011rb,Banfi:2012du,Catani:2015vma,Monni:2016ktx,Isaacson:2017hgb,Coradeschi:2017zzw,Kang:2017cjk,Scimemi:2017etj,Camarda:2019zyx},
reaching nowadays the standard of next-to-next-to-next-to-leading-logarithmic
(N$^3$LL) accuracy
\cite{Bizon:2017rah,Bizon:2018foh,Bizon:2019zgf,Bacchetta:2019sam,Becher:2019bnm,Ebert:2020dfc,Becher:2020ugp,Re:2021con,Ju:2021lah,Camarda:2021ict,Isaacson:2023iui},
with some next-to-N$^3$LL (i.e.~N$^4$LL) elements approximately encoded in some cases
\cite{Neumann:2022lft,Camarda:2023dqn,Campbell:2023lcy}. The inclusion of QED multiple emissions and virtual EW effects in the paradigm of analytic resummation has been considered more recently. Early work \cite{Cao:2004yy} and phenomenological studies \cite{Balossini:2009sa,CarloniCalame:2016ouw} were produced focusing on the impact of EW corrections on precision CCDY observables. Analytic ingredients for a mixed-coupling resummation were computed in \cite{deFlorian:2015ujt,deFlorian:2016gvk,Billis:2019evv}. A combination of
QCD, QED and mixed QCD-QED resummations was achieved in
\cite{Cieri:2018sfk,Autieri:2023xme} for on-shell $Z$ and $W$ production,
respectively, i.e.~without taking into account leptonic decay products. QED resummation effects at the level of final-state leptons have been so far available only through dedicated QED shower programs such as \pho{} \cite{Golonka:2005pn} and \horace{}~\cite{CarloniCalame:2003ux,CarloniCalame:2005vc}, including the possibility to match with fixed order NLO-EW results, or general-purpose Monte Carlo event generators as \py/\vin{}~\cite{Sjostrand:2014zea,Fischer:2016vfv,Kleiss:2017iir},
\hw{}~\cite{Bellm:2015jjp,Bewick:2023tfi} and \sherpa{}~\cite{Schonherr:2008av,Krauss:2018djz,Sherpa:2019gpd}. These tools, however, typically feature a quite low logarithmic accuracy, which may be a limiting factor for their use in high-precision phenomenology. In this context, the state of art is represented by matched calculations which include a combination of factorisable effects of both QCD and EW origin, and preserve the NLO-QCD and NLO-EW accuracy for inclusive quantities with respect to additional radiation~\cite{Bernaciak:2012hj,Barze:2012tt,Barze:2013fru,Chiesa:2024qzd,Muck:2016pko}.

In this article we take a step forward in the inclusion of EW
effects in the DY process. We present a highly accurate combination of QCD, EW, and mixed QCD-EW resummations for DY production at the level of final-state lepton pairs, derived in the \rad{} framework \cite{Monni:2016ktx,Bizon:2017rah,Re:2021con}. Our predictions include all necessary terms for a next-to-leading-logarithmic (NLL) resummation in the EW coupling constant $\aw$, as well as in the mixed QCD-EW expansion $\as\aw$, on top of retaining N$^3$LL QCD accuracy. First-order constant terms in the EW and in the mixed expansions, as well as third-order constant terms in QCD, are also included, paving the way to a state-of-the-art matching of resummed predictions with fixed-order calculations. The capability of describing final-state leptons is unavoidable to realistically match the setup of experimental DY analyses,
which feature fiducial acceptance cuts on the leptonic system. Our results allow one to assess at unprecedented accuracy the impact of all-order mixed QCD-EW corrections on leptonic observables, such as the charged-lepton transverse momentum, the lepton-pair transverse mass, or the jacobian asymmetry \cite{Rottoli:2023xdc,Torrielli:2023tiz}, relevant for $W$-mass determination.
Moreover, they open the door to the exploration of EW effects in different resummation environments, still available in \rad{}, such as jet-vetos or double-differential resummations \cite{Monni:2019yyr,Kallweit:2020gva}, or for other scattering processes.

The article is organised as follows. In section \ref{sec:formalism} we concisely
review the \rad{} resummmation formalism, and detail how EW and mixed QCD-EW
effects are consistently included at NLL accuracy. In section
\ref{sec:validation} we describe the validation of our results. Section
\ref{sec:pheno} collects our phenomenological predictions at the LHC, both for
neutral- and for charged-current DY. Finally, we
draw our conclusions in section \ref{sec:end}. Appendix \ref{sec:formulae}
collects formul\ae{} relevant for the employed theoretical framework.

\section{Inclusion of EW and mixed QCD-EW effects in \rad}
\label{sec:formalism}
The \rad{} formalism \cite{Monni:2016ktx,Bizon:2017rah,Re:2021con} is designed for the all-order resummation of enhanced logarithmic effects in global recursively infrared- and collinear- (rIRC) safe \cite{Banfi:2001bz,Banfi:2003je,Banfi:2004yd} observables that vanish away from the Sudakov limit. Notable observables in this class include for instance the transverse momentum of the final-state Drell Yan leptonic system, where the limit of small observable is determined \cite{Parisi:1979se} by azimuthal cancellations among the emitted radiation.
The formalism is based on a physical picture in which hard particles incoming to or outgoing from a primary scattering coherently radiate an ensemble of soft and collinear partons. The resummation is performed in momentum space, as opposed to conventional impact-parameter ($b$) space, namely the expressions are directly written in terms of the momenta of the radiated partons.

Radiative effects on rIRC-safe observables can be systematically classified according to the perturbative logarithmic order at which they enter. We denote with $V$ the considered observable, that we assume to be dimensionless without loss of generality. $\Sigma(v)$ represents the cumulative cross section for $V$ being smaller than $v$. In a gauge theory with coupling $a$, the counting of logarithms is performed at the level of $\ln\Sigma(v)$, where terms of order $a^n \ln(1/v)^{n+1-k}$ are ranked as (next-to)$^k$-leading logarithmic (N$^k$LL), $n$ and $k$ being positive integers.

Focusing first on the case of QCD, with $a=\as(\muR) \equiv \as$, the strong coupling at the renormalisation scale $\muR$, the \rad{} formula for the resummation of $V$ in colour-singlet production can be schematically written as
\beq
\label{eq:master-kt-space}
\frac{\rd\Sigma(v)}{\rd\Phi_B}
& = &
\int
\frac{\rd k_{t1}}{k_{t1}} \,
{\cal L}(k_{t1}) \, \re^{-R(k_{t1})} \, {\cal F}(v, \Phi_B, k_{t1})
\, ,
\eeq
where the expression is fully differential with respect to the Born phase-space variables $\Phi_B$, which allows for the application of fiducial cuts to match experimental acceptance.

The Sudakov radiator $R(k_{t1})$ is defined as
\beq
\label{eq:sud}
&&
R(k_{t1})
=
\sum_{\ell=1}^2
R_\ell(k_{t1})
\, ,
\qquad\quad
R_\ell(k_{t1})
\, = \,
\int_{k_{t1}}^M
\frac{\rd q}{q}
\Big[
A_\ell(\as(q)) \ln\frac{M^2}{q^2}
+
B_\ell(\as(q))
\Big]
\, ,
\eeq
where $k_{t1}$ denotes the transverse momentum of the hardest radiation in the ensemble of emitted QCD partons, while $M$ is the hard scale of the process, e.g.~the lepton-pair invariant mass for Drell Yan. $A_\ell$, $B_\ell$ are flavour-conserving soft-collinear  and hard-collinear anomalous dimensions with a well-defined perturbative expansion:
\beq
A_\ell(\as)
\, = \,
\sum_{n=1}^\infty
\Big(\frac\as{2\pi}\Big)^n
A_\ell^{(n)}
\, ,
\qquad\quad
B_\ell(\as)
\, = \,
\sum_{n=1}^\infty
\Big(\frac\as{2\pi}\Big)^n
B_\ell^{(n)}
\, .
\eeq
In the above formul\ae, $\ell$ labels the hard legs responsible for radiation, with $\ell = 1,2$ for the initial-state radiation relevant to colour-singlet production in QCD. The dependence upon the flavour of the hard legs is encoded in the values of the corresponding anomalous dimensions $A_\ell$ and $B_\ell$.
The evaluation of the integral in \eq{eq:sud} yields
\beq
\label{eq:sud2}
R(k_{t1})
\, = \,
- \, L \, g_1(\lambda) -
\sum_{n=0}^\infty
\Big(
\frac{\as}\pi
\Big)^n \, g_{n+2}(\lambda)
\, ,
\eeq
where $\lambda=\as \, \beta_0 \, L$, and $\beta_0$ is the first coefficient of the QCD beta function. Here $L=\ln(Q/k_{t1})$, with $Q$ being the resummation scale, namely a hard scale of the order of $M$, whose variations allow one to estimate the impact of missing higher-order logarithmic towers. Explicit expressions for the $g_{1,2}(\lambda)$ functions, as well as for the corresponding anomalous dimensions are collected in appendix \ref{sec:formulae}, while functions $g_{3,4}(\lambda)$ were presented in appendix B of \cite{Bizon:2017rah}.

The luminosity factor ${\cal L}(k_{t1})$ in \eq{eq:master-kt-space} incorporates the Born matrix element and PDF combination, as well as the hard virtual function $H(\muR)$ and collinear coefficient functions $C_{ab}(z)$:
\beq
\label{eq:lumi}
{\cal L}(k_{t1})
\, = \,
\sum_{c,d}
|{\cal M}_B|^2_{cd} \, \,
\sum_{i}
\Big[
C_{ci}
\otimes
f_i(k_{t1})
\Big](x_1)
\, \,
\sum_{j}
\Big[
C_{d\hspace{0.08mm}j}
\otimes
f_j(k_{t1})
\Big](x_2)
\, \,
H(\muR)
\, ,
\quad
\eeq
where the convolution is defined by $[f\otimes g](x) = \int_x^1\frac{\rd z}{z} \, f(z) \, g(x/z)$, and
\beq
\label{eq:CH_QCD}
C_{ab}(z)
& = &
\delta_{ab} \, \delta(1-z)
+
\sum_{n=1}^\infty
\Big(
\frac{\as}{2\pi}
\Big)^n \, C_{ab}^{(n)}(z)
\, ,
\nnb\\[4pt]
H(\muR)
& = &
1
+
\sum_{n=1}^\infty
\Big(
\frac{\as}{2\pi}
\Big)^n \, H^{(n)}(\muR)
\, .
\eeq
In \eq{eq:lumi} the sums run over all flavour combinations relevant to the considered process, $|{\cal M}_B|^2_{cd}$ is the Born squared matrix element, and $f_i(k_{t1})$ are the parton densities evaluated at scale $k_{t1}$.
As customary in the \rad{} approach, see e.g.~\cite{Bizon:2017rah}, the hard factor $H(\muR)$ in \eq{eq:CH_QCD} also includes constant contributions stemming from the Sudakov radiator, expanded out at the appropriate order in the coupling constants: these originate from the introduction of a resummation scale $Q\neq M$ in the definition of the resummed logarithm $L$, whence they induce an explicit dependence on $Q$ and $M$ in $H(\muR)$. Moreover, all factors of $\as(k_{t1})$ and $f_k(k_{t1})$ appearing in the luminosity ${\cal L}(k_{t1})$ are rewritten in terms of $\as(\muR \, \re^{-L})$ and $f_k(\muF \, \re^{-L})$, with $\muF$ being the factorisation scale, and then perturbatively expanded. As a consequence, the $C_{ab}(z)$ coefficient functions in \eq{eq:CH_QCD} acquire an explicit dependence on $Q$, $\muF$, and $\muR$. The evolution of PDFs between different scales is ruled by the DGLAP \cite{Altarelli:1977zs,Gribov:1972ri,Dokshitzer:1977sg} equation
\beq
\label{eq:DGLAP1}
\frac{\partial f_i(\mu,x)}{\partial \ln\mu}
\, = \,
\frac{\as(\mu)}{\pi}
\Big[
\hat P_{ij}
\otimes
f_j(\mu)
\Big](x)
\, ,
\qquad\quad
\hat P_{ij}(z)
\, = \,
\sum_{n=0}^\infty
\Big(
\frac{\as}{2\pi}
\Big)^n
\hat P^{(n)}_{ij}(z)
\, ,
\eeq
in terms of the regularised splitting functions $\hat P_{ij}$.

The last ingredient in \eq{eq:master-kt-space} is the radiative function ${\cal F}(v, \Phi_B, k_{t1})$, describing an arbitrary number of resolved soft and/or collinear real emissions with a transverse momentum smaller than $k_{t1}$, starting at NLL accuracy.
An explicit expression for this function is not relevant for the present discussion, and will not be derived here. It can however be extracted from the detailed construction of \cite{Bizon:2017rah,Re:2021con}. The interested reader can refer to formula (3.33) of Ref.~\cite{Re:2021con} for the evaluation of \eq{eq:master-kt-space} up to N$^3$LL$'$ order in QCD, using the ingredients calculated in \cite{Catani:2011kr,Catani:2012qa,Gehrmann:2014yya,Luebbert:2016itl,
Echevarria:2016scs,Li:2016ctv,Vladimirov:2016dll,Moch:2017uml,Moch:2018wjh,Lee:2019zop,Luo:2019bmw,Henn:2019swt,Bruser:2019auj,Henn:2019rmi,Luo:2019szz,vonManteuffel:2020vjv,Ebert:2020yqt,Luo:2020epw}. N$^3$LL$'$ accuracy gives control over all logarithmic towers up to $\as^n \ln(1/v)^{n-2}$, as well as all terms of order $\as^n \ln(1/v)^{2n-6}$ in the expanded cumulative cross section. We consider that equation as our baseline QCD resummation formula, and focus on the elements necessary to augment it with EW effects.\\

Logarithmically enhanced QED and mixed QCD-EW contributions stem from different sources. We aim at reaching NLL accuracy in the mixed coupling expansion, namely at correctly resumming all terms of order $\as^n \aw^m \ln(1/v)^{n+m}$, with $\aw=\aw(\muR)$ the QED running coupling evaluated at the renormalisation scale.

The first effect we consider is the QED contribution to the Sudakov radiator relevant to hard leg $\ell$. Analogously to \eq{eq:sud}, we have
\beq
\label{eq:EWsud0}
R^{\rm QED}_\ell(k_{t1})
\, = \,
\int_{k_{t1}}^M
\frac{\rd q}{q}
\, 
\Big[
A'_\ell(\aw(q)) \ln\frac{M^2}{q^2}
+
B'_\ell(\aw(q))
\Big]
\, ,
\eeq
where $A'_\ell$ and $B'_\ell$ are the abelian versions \cite{deFlorian:2015ujt,deFlorian:2016gvk} of the corresponding QCD anomalous dimensions:
\beq
\label{eq:EWAB}
A'_\ell(\aw)
\, = \,
\sum_{n=1}^\infty
\Big(\frac\aw{2\pi}\Big)^n
A_\ell^{\prime \, (n)}
\, ,
\qquad\quad
B'_\ell(\aw)
\, = \,
\sum_{n=1}^\infty
\Big(\frac\aw{2\pi}\Big)^n
B_\ell^{\prime \, (n)}
\, .
\eeq

Since the Drell Yan process features more than two charged particles at Born
level, soft wide-angle QED radiation introduces correlations among the hard
legs, namely it cannot be described as the incoherent sum of single-leg
contributions. This effect, starting at NLL QED accuracy (i.e.
$\aw^n \ln(1/v)^n$), is accounted for by including a radiative function
$D'(\aw(q),\Phi_B)$ in the Sudakov exponent
\cite{Catani:2014qha,Autieri:2023xme}, with a QED perturbative expansion: \beq
D'(\aw,\Phi_B) \, = \, \sum_{n=1}^\infty \Big(\frac\aw{2\pi}\Big)^n
D^{\prime(n)}(\Phi_B) \, . \eeq As the notation suggests, such a function
carries an explicit dependence upon the Born kinematics, through the invariant
masses of charge-correlated pairs. Its expression for \emph{massive} charged
leptons, such as the ones we consider throughout this article, can be deduced as
the abelian version of the corresponding QCD function relevant for heavy-quark
production \cite{Catani:2014qha}. We stress that a finite charged-lepton mass is
necessary in order to apply our resummation formalism as is. Explicit logarithms
of the mass are generated in the $D'(\aw(q),\Phi_B)$ soft contribution, see \eq{appA:anomalous_dimensions}. This
description applies to the physical case of \emph{bare} muons, namely not clustered with surrounding photon radiation. For the case of
electrons, a calorimetric definition in terms of \emph{dressed} leptons is more
appropriate from the experimental point of view. This would require an extension
of our formalism to resum a new kind of observable, such as the dressed-lepton
analogue of the $q_{T}$
imbalance~\cite{Sun:2015doa,Sun:2016kkh,Sun:2018icb,Chien:2019gyf,Buonocore:2021akg},
and is left for future work.

The pure QED correction to the Sudakov radiator is then \beq
\label{eq:EWsud}
R^{\rm QED}(k_{t1})
\, = \,
\int_{k_{t1}}^M
\frac{\rd q}{q}
\, 
\bigg\{
\sum_{\ell=1}^2
\Big[A'_\ell(\aw(q)) \ln\frac{M^2}{q^2}
+
B'_\ell(\aw(q))
\Big]
+
D'(\aw(q),\Phi_B)
\bigg\}
\, ,
\eeq
where $\ell$ ranges in $1,2$, as is the case for QCD, since no collinear singularities are associated to massive leptons, i.e.~the corresponding $A_\ell'$ and $B_\ell'$ functions are null. Up to NLL QED accuracy, \eq{eq:EWsud} can be cast as
\beq
R^{\rm QED}(k_{t1})
\, = \,
- \, L \, g_1'(\lambda') - \, g_2'(\lambda')
\, ,
\eeq
in terms of $\lambda' = \aw \, \beta_0' \, L$, with $\beta_0'$ being the first coefficient of the QED beta function. The $g_{1,2}'(\lambda')$ functions are written in terms of the anomalous dimensions $A_\ell^{\prime (1)}$, $A_\ell^{\prime (2)}$, $B_\ell^{\prime (1)}$, and $D^{\prime (1)}(\Phi_B)$, whose expressions are collected in appendix \ref{sec:formulae}.

Genuine mixed QCD-EW contributions to the Sudakov radiator stem from QED (QCD) corrections to QCD (QED) running couplings, as well as from mixed ${\cal O}(\as^n\aw^m)$ soft-collinear and hard-collinear anomalous dimensions, $A^{(n,m)}$ and $B^{(n,m)}$ respectively. We note that the ${\cal O}(\as\aw)$ soft-collinear anomalous dimension $A^{(1,1)}$ is null, hence at NLL in the mixed expansion (terms of order $\as^n \aw^m \ln(1/v)^{n+m}$) the correction to the radiator simply amounts to
\beq
R^{\rm MIX}(k_{t1})
\, = \,
- \,
\frac1{2\pi}
\sum_{\ell=1}^2
\int_{k_{t1}}^M
\frac{\rd q}{q}
\, 
\bigg[
\frac{\as^2 \, \beta_{01}\ln\xi'}{\xi^2\beta_0'}
A^{(1)}_\ell
+
\frac{\aw^2\beta'_{01}\ln\xi}{\xi^{\prime 2}\beta_0}
A^{\prime(1)}_\ell
\bigg]
\ln\frac{M^2}{q^2}
\, ,
\eeq
with $\xi=1-2 \, \as \, \beta_0 \ln\frac{\muR}q$, $\xi'=1-2 \, \aw \, \beta_0' \ln\frac{\muR}q$, and $\beta_{01}$ ($\beta_{01}'$) representing the lowest-order QED (QCD) contribution to the QCD (QED) running, see also \cite{Cieri:2018sfk,Billis:2019evv}. The result can be written as
\beq
R^{\rm MIX}(k_{t1})
\, = \,
- \, g_{11}(\lambda,\lambda')
- \, g'_{11}(\lambda,\lambda')
\, ,
\eeq
with constituent functions again given in appendix \ref{sec:formulae}.

Although the $B^{(1,1)}$ coefficient \cite{Cieri:2020ikq} enters at NNLL accuracy, as it generates terms of order $\as^n \aw^m \ln(1/v)^{n+m-1}$, we nevertheless include it in the Sudakov exponent to correctly account for all single-logarithmic contributions of order $\as \,\aw \,\ln(1/v)$. Our complete radiator including EW effects then reads
\beq
\label{eq:finalR}
R(k_{t1})
\, = \,
\Big[
R(k_{t1})
\Big]
_{\rm eq.\,(\ref{eq:sud2})}
\, + \,
R^{\rm QED}(k_{t1})
\, + \,
R^{\rm MIX}(k_{t1})
\, + \,
\frac\as{2\pi} \,
\frac\aw{2\pi} \,
B^{(1,1)} \, L
\, .
\eeq
We refer to NLL$_{\rm EW}$ accuracy when considering EW effects stemming from $R^{\rm QED}(k_{t1}) + R^{\rm MIX}(k_{t1})$ in \eq{eq:finalR}, and to nNLL$_{\rm MIX}$ accuracy when including $B^{(1,1)}$ as well. The nomenclature suggests that such a term is of mixed QCD-EW origin, and is part of the NNLL correction in the mixed coupling expansion.

Turning now to the analysis of luminosity factor in \eq{eq:lumi}, its leading EW corrections amount to the following replacements:
\beq
\label{eq:HCEW}
C_{ab}(z)
& = &
\Big[ C_{ab}(z) \Big]_{\rm eq.\,(\ref{eq:CH_QCD})}
+
\frac{\aw}{2\pi} \, C_{ab}^{\,\prime(1)}(z)
+
\frac{\as}{2\pi} \,
\frac{\aw}{2\pi} \, C_{ab}^{(1,1)}(z)
\, ,
\nnb\\[4pt]
H(\muR)
& = &
\Big[ H(\muR) \Big]_{\rm eq.\,(\ref{eq:CH_QCD})}
+
\frac{\aw}{2\pi} \, H^{\prime(1)}(\muR)
+
\frac{\aw}{2\pi} \, F^{\prime(1)}(\Phi_B)
+
\frac{\as}{2\pi} \, \frac{\aw}{2\pi}
\, H^{(1,1)}(\muR)
\, .
\eeq
In \eq{eq:HCEW}, $C_{ab}^{\,\prime(1)}(z)$ and $F^{\prime(1)}(\Phi_B)$ refer to ${\cal O}(\aw)$ QED constants of initial-state collinear and soft wide-angle origin, respectively,
obtained abelianising the corresponding QCD
expressions~\cite{deFlorian:2015ujt,deFlorian:2016gvk,Catani:2014qha}. $H^{\prime(1)}(\muR)$ is the EW one-loop virtual correction, that we evaluate with \recola{}~\cite{Actis:2016mpe,Denner:2017wsf}. The inclusion of primed quantities in \eq{eq:HCEW} allows one to reach NLL$'_{\rm EW}$ level, i.e.~to correctly capture all terms of order $\aw^n \ln(1/v)^{2n-2}$ in the
cumulative cross section. Quantities labelled with ``$(1,1)$'' in \eq{eq:HCEW}
formally enter at order $\as^n \aw^m \ln(1/v)^{n+m-2}$ in the cumulative cross
section, thus they are beyond NLL$'$ accuracy in both QCD and EW expansions.
However, they need to be included if one aims at matching the resummed
calculation with a fixed-order prediction at ${\cal O}(\as \aw)$ accuracy. We
define the accuracy attained by means of their inclusion as nNLL$_{\rm MIX}'$,
consistently with the nomenclature introduced above. Corresponding to the
modifications detailed in \eq{eq:HCEW}, DGLAP evolution is now ruled by \beq
\label{eq:PEW}
\hat P_{ij}(z)
\, = \,
\Big[ \hat P_{ij}(z) \Big]_{\rm eq.\,(\ref{eq:DGLAP1})}
+
\frac{\aw}{2\pi} \, \hat P_{ij}^{\,\prime(1)}(z)
+
\frac{\as}{2\pi} \, \frac{\aw}{2\pi} \, \hat P_{ij}^{(1,1)}(z)
\, ,
\eeq
in terms of the QED ($\hat P^{\,\prime(1)}$) and mixed QCD-QED ($\hat P^{(1,1)}$) splitting kernels reported in \cite{deFlorian:2015ujt,deFlorian:2016gvk}.

A concluding remark on the inclusion of photon-initiated contributions is in order. A photon PDF in the luminosity ${\cal L}(k_{t1})$ is needed in the context of EW corrections to Drell Yan. This is due to the presence of $C^{\,\prime}_{q\gamma}(z)$ coefficient functions in \eq{eq:HCEW}, as well as to QED contributions to DGLAP evolution in \eq{eq:PEW}.
Moreover, in the case of NCDY, a purely photon-induced Born channel $|{\cal M}_B|^2_{\gamma\gamma}$ is active. Although its impact on the fiducial cross section is at the percent level with respect to QCD corrections, see \cite{Bonciani:2021zzf}, its effects on differential distributions are not necessarily negligible with respect to the other EW corrections we include. In our simulations we consider all photon contributions mentioned above, and consistently adopt PDF sets that feature a photon density \cite{Manohar:2016nzj}. We instead refrain from including photon-initiated channels in the ${\cal O}(\as\aw)$ constant contribution, as numerically negligible \cite{Bonciani:2021zzf}.

\section{Validation}
\label{sec:validation}
In this section we provide a validation of our implementation of EW effects in Drell Yan. We start by describing the physical setup we employ. We consider NCDY at the LHC, $pp\to Z/\gamma^* \, (\to \mu^+\mu^-) +X$, with centre-of-mass energy $\sqrt s= 14$ TeV. For the EW couplings we use the $G_\mu$ scheme
with $G_F = 1.1663787\times10^{-5}$ GeV$^{-2}$ and set on-shell masses and widths to the values $m_{W,\rm OS} = 80.385$ GeV, $m_{Z,\rm OS} = 91.1876$ GeV, $\Gamma_{W,\rm OS} = 2.085$ GeV, and $\Gamma_{Z,\rm OS} = 2.4952$ GeV. Such mass values are converted to pole masses via the formula $m_V = m_{V,\rm OS} \, (1+\Gamma_{V,\rm OS}^2/m_{V,\rm OS}^2 )^{-1/2}$, with $V=W,Z$. The EW coupling is determined as $\aw=\sqrt2 \, G_F \, m_W^2 \, (1-m_W^2/m_Z^2)/\pi$, and the complex-mass scheme \cite{Denner:2005fg} is employed throughout. We consider massive muons, with $m_\mu = 105.658369$ MeV. Higgs and top-quark pole masses are set to $m_H = 125.9$ GeV and 
$m_t = 173.07$ GeV, respectively. We use a diagonal CKM matrix. We assume $n_f = 5$ active quark flavours, and retain the exact $m_t$ dependence in all virtual and real-virtual amplitudes associated to bottom-induced processes, except for the two-loop virtual corrections, where top-mass effects are neglected.
We use the \texttt{NNPDF31\_nnlo\_as\_0118\_luxqed} PDF set \cite{Bertone:2017bme}, which is based on the LUXqed methodology \cite{Manohar:2016nzj} for the determination of the photon content of the proton.  PDF sets are included through the \texttt{LHAPDF} interface \cite{Buckley:2014ana}. DGLAP evolution, including the photon PDF, as well as convolutions with coefficient functions are performed by means of the \texttt{Hoppet} package \cite{Salam:2008qg}. All fixed-order predictions presented in the following, including those used for matching, are obtained with the \matr{} code~\cite{Grazzini:2017mhc}. More precisely, mixed QCD-EW corrections are validated against the calculations of Refs.~\cite{Buonocore:2021rxx,Bonciani:2021zzf}\footnote{We thank the authors of Refs.~\cite{Buonocore:2021rxx,Bonciani:2021zzf} for providing us with a private version of \matr{} to perform the validation at ${\cal O}(\as\aw)$.}. Renormalisation and factorisation scales are set to $\muR = \muF = m^{\mu\mu}$, the di-muon invariant mass. To ensure a consistent matching between the \rad{} and the \matr{} predictions, we set $\aw(\muR)=\aw|_{G_\mu}$, i.e.~independent of the value of $\mu_R$. The following selection cuts on the leptonic system are applied:
\beq
\label{eq:cuts1}
p_t^{\mu^\pm} > 25 \, {\rm GeV} \, , \qquad |y^{\mu^\pm}| < 2.5 \, , \qquad
m^{\mu\mu} > 50 \, {\rm GeV} \, , \eeq with $p_t^{\mu^\pm}$ and $y^{\mu^\pm}$
the transverse momentum and rapidity of muons. Muons are considered at the bare
(as opposed to dressed) level. The two-loop ${\cal O}(\as\aw)$ corrections are calculated in
the pole
approximation~\cite{Dittmaier:2015rxo,Denner:2019vbn,Buonocore:2021rxx,Bonciani:2021zzf,Dittmaier:2024row}.
This is based on a systematic expansion of the cross section around the
heavy-boson resonance, in such a way that the radiative corrections are
separated into well-defined, gauge-invariant contributions.

In order to detail our validation procedure, we introduce an additive matching of the resummation with the fixed-order prediction:
\beq
\label{eq:add_match1}
\frac{{\rd}\sigma_{\rm RES+FO}}{{\rd}p_t^{\mu\mu}}
\, = \,
\frac{{\rd}\sigma_{\rm RES}}{{\rd}p_t^{\mu\mu}}
\, + \,
\frac{{\rd}\sigma_{\rm FO}}{{\rd}p_t^{\mu\mu}}
\, - \,
\Big[
\frac{{\rd}\sigma_{\rm RES}}{{\rd}p_t^{\mu\mu}}
\Big]_{\rm FO}
\, .
\eeq
All contributions to the previous equation are differential cross sections with respect to the di-muon transverse momentum $p_t^{\mu\mu}$, as well as to all Born degrees of freedom $\Phi_B$.
The ${\rd}\sigma_{\rm RES}$ term represents the resummed cross section detailed in Sec.~\ref{sec:formalism}. In our case, the resummation is evaluated at NLL$'_{\rm EW}$ (nNLL$_{\rm MIX}'$) accuracy upon excluding (including) the terms labelled with ``(1,1)'' in eqs.~(\ref{eq:finalR}) to (\ref{eq:PEW}). The ${\rd}\sigma_{\rm FO}$ component is the fixed-order calculation for the DY process in presence of resolved radiation. Corresponding to a resummation at NLL$'_{\rm EW}$ (nNLL$_{\rm MIX}'$) accuracy, it includes corrections up to ${\cal O}(\aw)$ (${\cal O}(\as\aw)$) with respect to Born level. Finally, $\big[{\rd}\sigma_{\rm RES}\big]_{\rm FO}$ is the perturbative expansion of the resummed contribution ${\rd}\sigma_{\rm RES}$, retaining the same order as featuring in ${\rd}\sigma_{\rm FO}$, which removes the overlap between the two latter contributions.

Provided ${\rd}\sigma_{\rm RES}$ in \eq{eq:add_match1} does \emph{not} contain a resummation of subleading-power corrections through transverse-recoil effects \cite{Catani:2015vma,Ebert:2020dfc}, the inclusive $p_t^{\mu\mu}$ integration of \eq{eq:add_match1} yields the fixed-order cross section (differential in $\Phi_B$) according to the $q_T$-subtraction formalism \cite{Catani:2007vq}.
The main technical challenge in the implementation of \eq{eq:add_match1} is related to the fact that ${\rd}\sigma_{\rm FO}$ and $\big[{\rd}\sigma_{\rm RES}\big]_{\rm FO}$ are separately divergent in the small-$p_t^{\mu\mu}$ limit, and only their difference is integrable. This is typically handled by introducing a slicing parameter $r_{\rm cut}$ (or $p_{t,\rm cut}^{\mu\mu}$) and cutting off such a difference for $p_t^{\mu\mu}/m^{\mu\mu} < r_{\rm cut}$ (or $p_t^{\mu\mu} < p_{t,\rm cut}^{\mu\mu}$). The correct fixed-order rate is obtained ideally by taking the limit of slicing parameter going to 0, in practice by considering as small cut-off values as possible, compatibly with the numerical stability of the result.

\begin{figure}[htpb]
\centering
\includegraphics[width=0.45\linewidth]{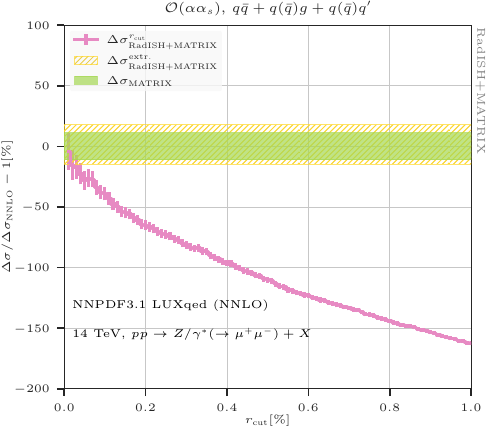}
\qquad
\includegraphics[width=0.45\linewidth]{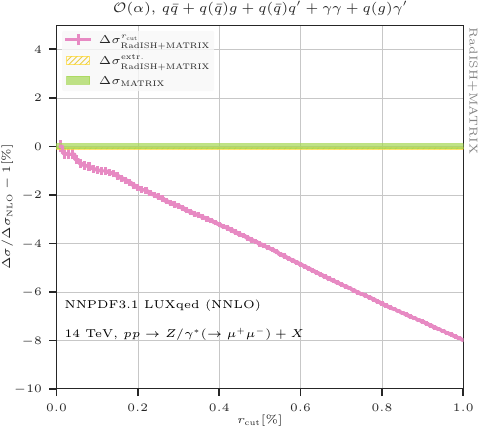}
\caption{Fixed-order validation of the ${\cal O}(\as\aw)$ (left panel) and ${\cal O}(\aw)$ (right panel) contributions to the fiducial cross section.}
\label{fig:validation1}
\end{figure}

In Fig.~\ref{fig:validation1} we validate our implementation of \eq{eq:add_match1} at the level of fiducial cross section (i.e.~inclusively integrated over $p_t^{\mu\mu}$), separately for the ${\cal O}(\as\aw)$ contribution (left panel), and for the ${\cal O}(\aw)$ contribution (right panel).
The displayed results are summed over all contributing partonic channels, but validation plots of similar quality (except for the reduced statistics) have been produced for the individual channels. The pink bars are \rad{}+\matr{} predictions as functions of $r_{\rm cut}$, with bar widths representing the numerical integration error associated with the result. 
The \rad{}+\matr{} label indicates that \rad{} is responsible for the resummation components  of \eq{eq:add_match1}, while \matr{} provides the fixed order. The results labelled as \matr{} are based instead on an independent implementation of $q_T$ subtraction. For consistency, the same fixed order component is used for the two predictions. The yellow band is an analytic extrapolation of the \rad{}+\matr{} result to $r_{\rm cut}\to 0$, obtained by fitting the pink curve with a linear $p_t^{\mu\mu}$ function enhanced by logarithms of $p_t^{\mu\mu}$. The green reference band is the fixed-order correction to the fiducial cross section as obtained with \matr{}, and results are displayed as a relative difference with respect to the latter. We note that the size of the extrapolation band generally depends on the specific function used for the fit of the $r_{\rm cut}$ dependence. The functions used in the \rad{}+\matr{} predictions include the expected powers of logarithmically-enhanced contributions at ${\cal O}(\as\aw)$ and ${\cal O}(\aw)$, while \matr{} always adopts a quadratic function for its fit.

Inspection of the two panels of Fig.~\ref{fig:validation1} immediately reveals the computational challenges related to these calculations: large linear power corrections in $p_t^{\mu\mu}$ require extremely small values of $r_{\rm cut}$ for the $q_T$-subtracted prediction to become asymptotic, especially for the ${\cal O}(\as\aw)$ correction. Such logarithmically-enhanced linear power corrections cannot be entirely reabsorbed via transverse recoil \cite{Catani:2015vma,Ebert:2020dfc}, for instance using the procedure outlined in \cite{Buonocore:2021tke,Camarda:2021jsw}, as they are in part caused by EW radiation \cite{Buonocore:2019puv} off the final-state leptons.
For all coupling combinations we consider, the \rad{}+\matr{} prediction correctly reproduces the \matr{} result in the asymptotic \mbox{$r_{\rm cut}\to0$} limit, within the respective numerical uncertainties. This represents a particularly powerful test for all aspects of the implementation. In particular, the logarithmic structure of the expanded cross section $\big[{\rd}\sigma_{\rm RES}\big]_{\rm FO}$ is checked with high accuracy to reproduce the one of the fixed-order calculation ${\rd}\sigma_{\rm FO}$, which is based on an independent numerical implementation. Moreover, a positive outcome of the plots in Fig.~\ref{fig:validation1} also tests that the cumulative resummed prediction and its perturbative expansion coincide asymptotically in the $p_t^{\mu\mu}\to\infty$ limit. Although for the sake of clarity we show this behaviour only for $\muR=\muF=m^{\mu\mu}$ in Fig.~\ref{fig:validation1}, we have successfully tested it for all 7 uncorrelated $\muR$ and $\muF$ variations around the central choice.

\begin{figure}[htpb]
\centering
\includegraphics[width=0.45\linewidth]{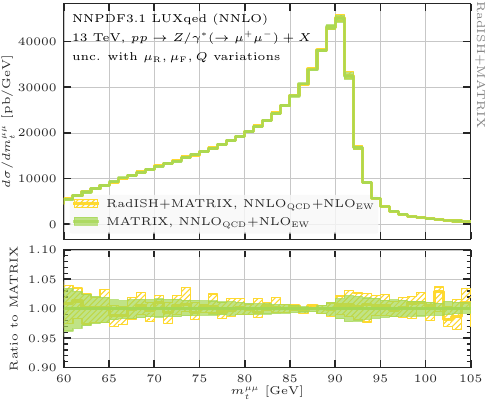}
\caption{Fixed-order comparison of the \rad{}+\matr{} implementation against \matr{}, differentially with respect to the di-muon transverse mass.}
\label{fig:validation2}
\end{figure}

In Fig.~\ref{fig:validation2} we show the di-muon transverse-mass distribution at fixed order, including NNLO QCD (i.e.~${\cal O}(\as^2)$) and NLO EW (i.e.~${\cal O}(\aw)$) corrections with respect to Born level, obtained both with \rad{}+\matr{} (yellow) and with \matr{} (green). The transverse mass is defined as $m_t^{\mu\mu}=\big(2 \, p_t^{\mu^+}p_t^{\mu^-}(1-\cos\Delta\phi^{\mu\mu})\big)^{1/2}$, with $\Delta\phi^{\mu\mu}$ being the azimuthal separation of the two leptons. The plot is obtained with a slicing parameter $p_{t,\rm cut}^{\mu\mu} = 0.1$ GeV. It employs the same setup as detailed above, with the exception of the centre-of-mass energy, now set to $\sqrt s= 13$ TeV, and the the selection cuts of \eq{eq:cuts1}, which are replaced by the ATLAS cuts of Ref.~\cite{ATLAS:2019zci}:
\beq
\label{eq:cuts2}
p_t^{\mu^\pm} > 27 \, {\rm GeV}
\, ,
\qquad
|\eta^{\mu\pm}| < 2.5
\, ,
\qquad
66 \, {\rm GeV} < m^{\mu\mu} < 116 \, {\rm GeV}
\, .
\eeq
In the whole $m_t^{\mu\mu}$ phase space, the \rad{}+\matr{} fixed-order prediction is checked to precisely reproduce the \matr{} one both in shape and in normalisation. This holds for the central value of the prediction, as well as for the theoretical uncertainty band, obtained with a 7-point variation of $\muR$ and $\muF$ by factors of 2 around the common central value $m^{\mu\mu}$. The quality of the agreement is comparable across the entire $m_t^{\mu\mu}$ spectrum, namely both below and above the jacobian peak at $m_t^{\mu\mu}=m_Z$. As QCD and EW mechanisms have different relative importance in the various $m_t^{\mu\mu}$ regions, the successful comparison shown in Fig.~\ref{fig:validation2} is a highly non-trivial test of their correct inclusion within our numerical framework.
We stress that such a stringent differential test is possible only upon controlling the final-state leptons fully exclusively over their fiducial phase space, as our formalism in \eq{eq:master-kt-space} allows us to do.

\section{Phenomenological results}
\label{sec:pheno}

For the phenomenological results of this section we consider both NCDY and CCDY at the 13 TeV LHC, in Sec.~\ref{sec:NCDY} and Sec.~\ref{sec:CCDY} respectively.
 The setup we use for NCDY is detailed at the beginning of Sec.~\ref{sec:validation}, with the fiducial cuts of \eq{eq:cuts2}. In the case of CCDY, we consider the process $pp\to W^+ \, (\to \mu^+\nu_\mu) +X$. Our choice for central $\muR$ and $\muF$ scales is $\big((m^{\mu\nu})^2 + (p_t^{\mu\nu})^2\big)^{1/2}$, with $m^{\mu\nu}$ ($p_t^{\mu\nu}$) the muon-neutrino invariant mass (transverse momentum). In the resummed calculation this expression is approximated with $m^{\mu\nu}$, which is correct up to quadratic power corrections in $p_t^{\mu\nu}$. The fiducial volume is defined by the following cuts on the charged lepton:
\beq
26\,{\rm GeV} < p_t^{\mu^+} < 55\,{\rm GeV}
\, ,
\qquad
|\eta^{\mu+}| < 2.4
\, ,
\qquad
m_t^{\mu\nu}
> 40 \, {\rm GeV}
\, ,
\eeq
with $m_t^{\mu\nu}=\big(2 \, p_t^{\mu^+}p_t^{\nu}(1-\cos\Delta\phi^{\mu\nu})\big)^{1/2}$ the muon-neutrino transverse mass, and $\Delta\phi^{\mu\nu}$ the azimuthal separation between muon and neutrino.

All resummed predictions are obtained as detailed in Sec.~\ref{sec:formalism}, with a modified version of the resummed logarithms~\cite{Bozzi:2005wk,Banfi:2012yh,Bizon:2017rah} $L=\ln\big[(Q/k_{t1})^p+1\big]/p$, with $p=6$, in order to smoothly turn off logarithmic terms at $k_{t1}\gg Q$. The use of such logarithms, and consequently of a jacobian $J(k_{t1})={\rd}L/{\rd}\ln(Q/k_{t1})$ in the integration measure of \eq{eq:master-kt-space}, induces a controlled set of $p$-dependent subleading power corrections in the formalism. Such terms do not affect the logarithmic accuracy of the calculation and, after matching, they exactly cancel up to the accuracy of the fixed-order component. Our results are obtained with the inclusion of transverse-recoil effects, which allow for the resummation of linear power corrections due to initial-state radiation \cite{Catani:2015vma,Ebert:2020dfc,Re:2021con}.

We adopt the additive matching introduced in \eq{eq:add_match1} for all observables, except for the di-lepton transverse momentum $p_t^{\ell\ell}$ (equal to the di-muon transverse momentum $p_t^{\mu\mu}$ in NCDY, and to the muon-neutrino transverse momentum $p_t^{\mu\nu}$ in CCDY), where we consider a more general prescription \cite{Bizon:2017rah,Re:2021con}:
\beq
\label{eq:add_match2}
\frac{{\rd}\sigma_{\rm RES+FO}}{{\rd}p_t^{\ell\ell}}
\, = \,
\frac{{\rd}\sigma_{\rm FO}}{{\rd}p_t^{\ell\ell}}
\, + \,
Z(p_t^{\ell\ell}) \,
\bigg\{
\frac{{\rd}\sigma_{\rm RES}}{{\rd}p_t^{\ell\ell}}
\, - \,
\Big[
\frac{{\rd}\sigma_{\rm RES}}{{\rd}p_t^{\ell\ell}}
\Big]_{\rm FO}
\bigg\}
\, ,
\eeq
with
\beq
Z(p_t^{\ell\ell})
\, = \,
\Big[
1-\big(p_t^{\ell\ell}/p_{t0}\big)^u
\Big]^h
\Theta(p_{t0}-p_t^{\ell\ell})
\, ,
\eeq
and $u=2$, $h=3$. The dampening profile $Z(p_t^{\ell\ell})$ ensures a smooth transition around the $p_{t0}$ scale, from a soft regime $p_t^{\ell\ell}\ll p_{t0}$ in which the matched result must reduce to the resummed component, to a hard region $p_t^{\ell\ell}\gg p_{t0}$ where one must recover the fixed order. A choice of $Z(p_t^{\ell\ell})\neq 1$ alters the unitarity of the matching, namely the $p_t^{\ell\ell}$ integral of \eq{eq:add_match2} does not reproduce the fixed-order result, hence it cannot be employed for $q_T$ subtraction. Its use is however physically motivated for the $p_t^{\ell\ell}$ distribution, since the presence of the transition scale $p_{t0}$ gives an extra handle to assess the matching systematics affecting the prediction. In the results presented below we vary $p_{t0}$ in the range $\{2/3,1,3/2\}\times m_V$, with $m_V=m_Z$ ($m_W$) in NCDY (CCDY). The total uncertainty band assigned to matched predictions for the $p_t^{\ell\ell}$ distribution is the envelope of $9\times3=27$ variations, where 3 is the number of chosen values for $p_{t0}$, while 9 is the combination of the canonical 7 variations of $\muR$, $\muF$ at central resummation scale $Q=m^{\ell\ell}/2$, with $2$ variations of $Q$ at central $\muR = \muF$.

In the following analysis our main focus is on the perturbative features of the new EW corrections, hence we don't include in our predictions a model for non-perturbative QCD corrections.
Moreover, since currently there is no public implementation of the results of \cite{Buonocore:2021rxx,Bonciani:2021zzf,Buccioni:2022kgy}, the nNLL$^\prime_{\rm MIX}$ predictions we present in the next sections do not contain mixed QCD-EW corrections to the fixed-order ${\cal O}(\as\aw)$ component.
We point out that, working at the level of bare muons, the inclusion of these terms can have a non-negligible impact on physical distributions such as the charged-lepton transverse momentum \cite{Dittmaier:2024row}, since large logarithms of the muon mass enter the non-singular component $\rd\sigma_{\rm FO} - \big[\rd\sigma_{\rm RES}\big]_{\rm FO}$.
We leave for future work a complete matching at this order, necessary for a thorough comparison with LHC data.

\subsection{Neutral-current Drell Yan}
\label{sec:NCDY}

\begin{figure}[htpb]
\centering
\includegraphics[width=0.45\linewidth]{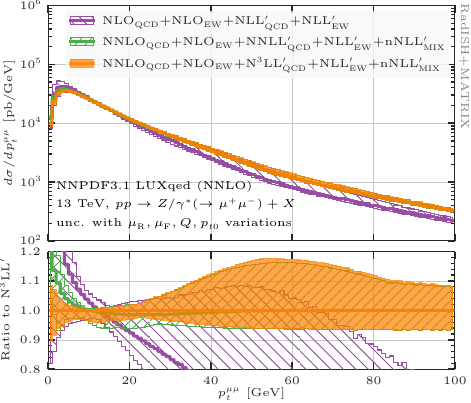}
\qquad
\includegraphics[width=0.45\linewidth]{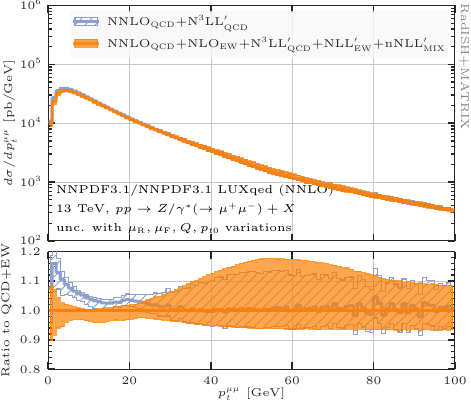}
\caption{Matched spectra for the di-lepton transverse momentum in neutral-current DY. Left panel: perturbative progression including QCD and EW effects. Right panel: effect of EW corrections on top of the QCD baseline.}
\label{fig:ZptB}
\end{figure}

We start by displaying in Fig.~\ref{fig:ZptB} the transverse momentum $p_t^{\mu\mu}$ of the di-muon system in NCDY. In the left panel we compare matched predictions with different accuracy. The purple band features NLO+NLL$'$ accuracy both in the QCD and in the EW coupling. We recall that this amounts to excluding all quantities with label ``(1,1)'' from eqs.~(\ref{eq:finalR}) to (\ref{eq:PEW}). Green and orange bands both include nNLL$'_{\rm MIX}$ EW effects (i.e. ``(1,1)'' quantities in eqs.~(\ref{eq:finalR}) to (\ref{eq:PEW})), as well as NNLO$_{\rm QCD}$, with the orange (green) attaining N$^3$LL$'$ (NNLL$'$) logarithmic QCD accuracy. At medium-large $p_t^{\mu\mu}$ the inclusion of NNLO$_{\rm QCD}$ contributions has the effect of significantly hardening the tail, and reducing the uncertainty band to the 10-15\% level. In the $p_t^{\mu\mu}\to0$ resummation region, nNLL$'_{\rm MIX}$ and especially NNLL$'_{\rm QCD}$ logarithmic terms lower the spectrum (green vs purple), a trend which is maintained after inclusion of N$^3$LL$'_{\rm QCD}$ contributions (orange vs green). We notice that in this region the uncertainty band is significantly reduced upon adding logarithmic effects, down to the few-\% level below 20 GeV for our most accurate prediction (orange). Predictions with higher formal accuracy are well contained within the uncertainty bands of lower orders in that region, which is a sign of good perturbative convergence.

In the right panel of Fig.~\ref{fig:ZptB} we assess the importance of including EW effects (orange) on top of the QCD NNLO+N$^3$LL$'$ baseline (light blue). The orange band is identical to the one in the left panel, which will be the case as well for the next figures in this section. The two predictions differ by their perturbative content, as well as by the PDF adopted, where a LUXqed photon PDF (together with its DGLAP evolution) is active only for the former. EW effects induce a visible distortion in the spectrum at small $p_t^{\mu\mu}$, lowering the prediction by as much as 10-15\% for $p_t^{\mu\mu} \lesssim 10$ GeV. We have checked that, as one might expect, EW corrections largely factorise from QCD in the small-$p_t^{\mu\mu}$ region, namely similar shape distortions as those in the right panel of Fig.~\ref{fig:ZptB} can be observed when including EW effects on top of lower-order QCD predictions. The same considerations apply for all observables considered below.
We also note that at small $p_t^{\mu\mu}$ the uncertainty bands of the two predictions are comparatively small, at the level of few \%, and do not overlap. The latter feature is not surprising, since EW corrections are genuinely new physical effects, whose magnitude is not supposed to be meaningfully estimated by pure-QCD scale variations. This consideration highlights the relevance of an accurate description of EW effects in DY production for a successful precision-physics programme at the LHC. The effect of all-order EW corrections becomes more and more marginal for $p_t^{\mu\mu} \gtrsim 30$ GeV (except for a slight increase in the uncertainty band in the matching region), where the prediction starts being dominated by the fixed-order component. In this region one also expects that the inclusion of non-factorisable ${\cal O}(\as\aw)$ QCD-EW effects, not considered in our results, may play a role.

\begin{figure}[htpb]
\centering
\includegraphics[width=0.45\linewidth]{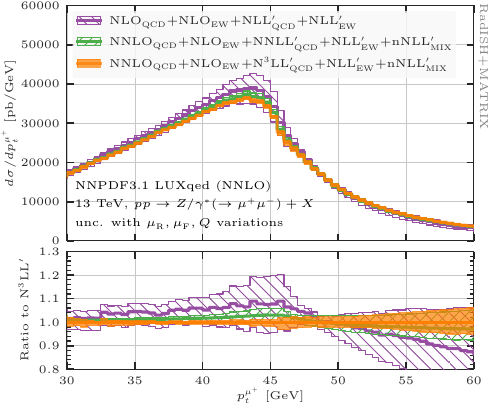}
\qquad
\includegraphics[width=0.45\linewidth]{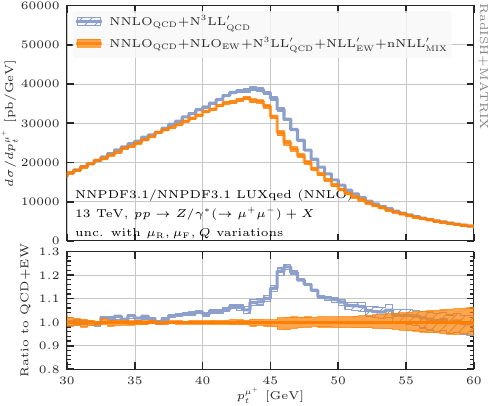}
\caption{Matched spectra for the positively charged muon transverse momentum in neutral-current DY. Left panel: perturbative progression including QCD and EW effects. Right panel: effect of EW corrections on top of the QCD baseline.}
\label{fig:Zptlep}
\end{figure}

In Fig.~\ref{fig:Zptlep} we display differential predictions with respect to the transverse momentum $p_t^{\mu^+}$ of the positively charged muon. The inclusion of resummation effects is necessary to provide a physical description of this observable~\cite{Catani:1997xc} due to its sensitivity to soft radiation for $p_t^{\mu^+}\simeq m^{\mu\mu}/2$. The pattern of the figure is identical to that of Fig.~\ref{fig:ZptB}, with the perturbative progression displayed in the left panel, and the impact of EW effects in the right panel. At variance with the di-muon transverse momentum, the $p_t^{\mu^+}$ spectrum is non-trivial already at Born level, hence we expect relatively milder perturbative corrections, and a solid perturbative stability across its entire phase space. This is what we find inspecting the left panel. Increasing QCD and EW formal accuracy (green vs purple) amounts to marginally lowering the jacobian peak and raising the tail at the level of roughly 5\%. The inclusion of yet higher-order QCD resummation continues the trend, with a further few-\% distortion. Theoretical uncertainty bands are found to reliably cover the central predictions of the next perturbative orders, both below and above the peak. The upgrade in formal accuracy has the visible effect of reducing the residual uncertainty, down to the level of $\pm2\%$ ($\pm4\%$) below (above) peak. As stated above, we expect however that a matching at ${\cal O}(\as\aw)$, not included in our predictions, will have a numerical impact on the $p_t^{\mu^+}$ distribution. This may exceed the quoted perturbative uncertainty, especially around the jacobian peak, due to genuine mixed effects which are not captured by scale variations.

The right panel of Fig.~\ref{fig:Zptlep} shows how the jacobian peak in $p_t^{\mu^+}$ is exposed to the interplay of QCD and EW effects. Including the latter has a clearly visible impact on the distribution, lowering the spectrum by as much as 20\% at $p_t^{\mu^+}\simeq m_Z/2$, in a way that by no means can be approximated by a constant rescaling factor. The shape of the correction is compatible with what observed in \cite{Alioli:2016fum} (see Fig.~24) in the context of a comparative study among event generators with QED resummation. In our case, the prediction including EW effects lies outside of the pure-QCD uncertainty band in the whole peak region, roughly from 35 GeV to 55 GeV. This accentuates what was observed in the right panel of Fig.~\ref{fig:ZptB} at small $p_t^{\mu\mu}$, highlighting the need for EW corrections for a complete description of this observable.

\begin{figure}[htpb]
\centering
\includegraphics[width=0.45\linewidth]{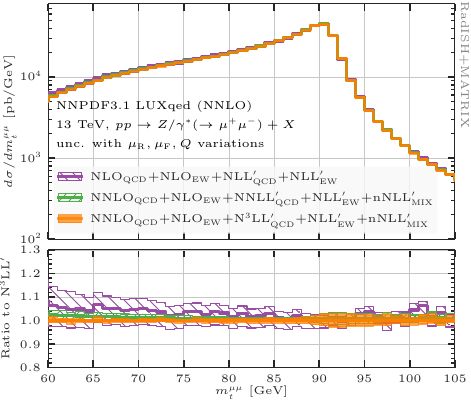}
\qquad
\includegraphics[width=0.45\linewidth]{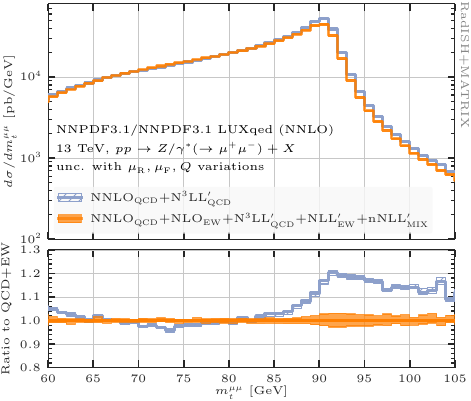}
\caption{Matched spectra for the di-muon transverse mass in neutral-current DY. Left panel: perturbative progression including QCD and EW effects. Right panel: effect of EW corrections on top of the QCD baseline.}
\label{fig:ZmTr}
\end{figure}

The di-muon transverse mass $m_t^{\mu\mu}$, displayed in Fig.~\ref{fig:ZmTr}, follows a similar pattern as the muon transverse momentum in Fig.~\ref{fig:Zptlep}. A solid perturbative convergence is observed in the left panel, both below and especially above the jacobian peak at $m_t^{\mu\mu}\simeq m_Z$. Perturbative corrections are relatively flat upon including EW effects, at the level of up to 5\% comparing purple and orange predictions. Uncertainty bands are significantly shrunk by the inclusion of subleading perturbative effects, again reaching $\pm2\%$ ($\pm4\%$) below (above) peak. The right panel shows that EW effects have moderate impact below the transverse-mass peak, with shape distortions at the $\pm3\%$ level for $m_t^{\mu\mu}\lesssim 85$ GeV. In the peak region and in the high-$m_t^{\mu\mu}$ tail the distortion reaches the 15-20\% level, with EW contributions consistently lowering the prediction.\\

\begin{figure}[htpb]
\centering
\includegraphics[width=0.44\linewidth]{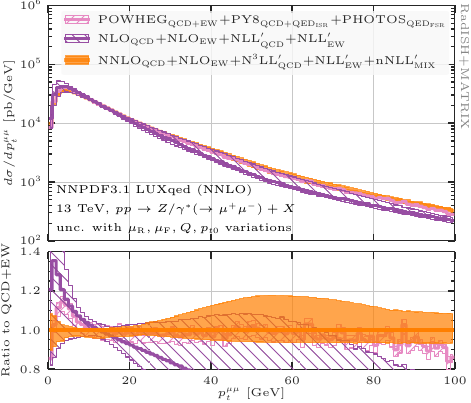}\qquad
\includegraphics[width=0.45\linewidth]{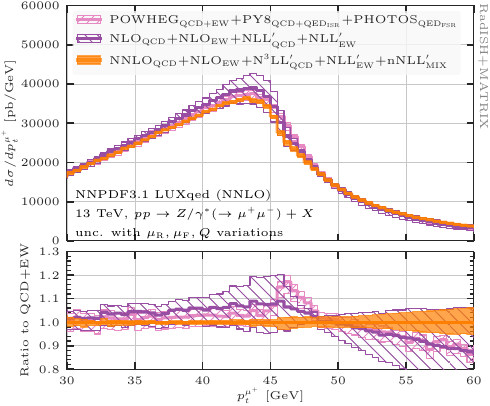}
\includegraphics[width=0.45\linewidth]{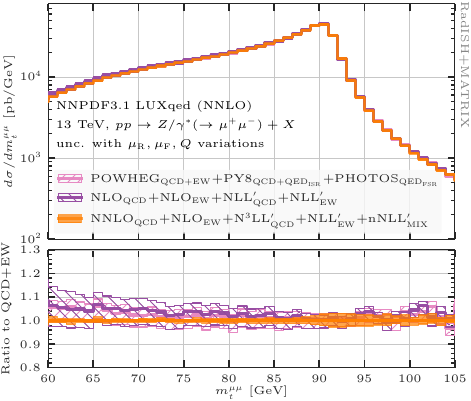}
\caption{Comparison of matched \rad{}+\matr{} spectra (purple and orange) against \pwg{} predictions (pink) for the di-muon transverse momentum, the positively charged muon transverse momentum, and the di-muon transverse mass in neutral-current Drell Yan.}
\label{fig:Z_vs_PWG}
\end{figure}

In Fig.~\ref{fig:Z_vs_PWG} we show the same observables that were considered in
Figs.~\ref{fig:ZptB} to \ref{fig:ZmTr}, comparing \rad{}+\matr{} predictions
against \pwg{} \cite{Barze:2012tt,Barze:2013fru} results. The latter tool
performs an NLO + parton shower (PS) matching including NLO QCD and NLO EW
effects at the level of matrix elements, as well as the resummation of QED and
QCD initial-state radiation (ISR) by means of \py \cite{Sjostrand:2014zea} (version 8.245), and the resummation of QED final-state radiation (FSR) by means of \pho{} \cite{Golonka:2005pn}. In order for the
comparison with \rad{}+\matr{} to be sensible, we do not consider hadronisation
and multi-particle interactions at the end of the \py{} showering phase. We
adopt the AZNLO tune~\cite{ATLAS:2014alx}, that was fit to
precise Drell-Yan $p_{t}^{\ell\ell}$ and $\phi_{\eta}^{*}$ data.
Moreover, we activate the \pwg{} flag {\tt lepaslight=0}, in order to treat the final-state muons as massive. \pwg{} contains
QCD and EW ingredients entering our NLO+NLL$^\prime$ results\footnote{We note that the photon-induced process $\gamma\gamma \to \mu^{+}\mu^{-}$ at LO is not available in the current version of the NCDY \pwg{} generator {\tt Z\_ew-BMNNPV} revision 4056.}. As such, \pwg{} predictions (pink curves in
Fig.~\ref{fig:Z_vs_PWG}) are expected to be fairly compatible with the \rad{}+\matr{} ones at NLO+NLL$^\prime$ accuracy (purple lines) within their respective uncertainties. Both are confronted to our best predictions (orange lines) to assess the numerical impact, with respect to the current state of the art, of the terms included in the present article for the first time. For clarity, we stress that the purple and orange
\rad{}+\matr{} predictions are the same (with identical colour code) as
displayed in the left panels of Figs.~\ref{fig:ZptB} to \ref{fig:ZmTr}.

Starting with the di-muon transverse momentum $p_t^{\mu\mu}$ in the upper-left panel of Fig.~\ref{fig:Z_vs_PWG}, we note that the \rad{}+\matr{} (purple) and \pwg{} (pink) central predictions are in reasonable shape agreement in the resummation region $p_t^{\mu\mu}\lesssim 20$ GeV. As far as the hard $p_t^{\mu\mu}$ tail is concerned, we instead observe a different shape between the two generators.
We have checked that the \rad{}+\matr{} result reproduces the fixed-order one
from $p_t^{\mu\mu}\simeq 50$ GeV on. Conversely, the transition region between
resummed and fixed-order regimes is shifted to larger transverse momenta and is broader in the \pwg{} description. This behaviour is controlled by the parameters ruling the exponentiation of non-singular contributions in the \pwgg{} Sudakov form factor \cite{Nason:2004rx,Alioli:2010xd}, implemented through the \pwgg{} damping mechanism.
The main criterion used to damp the non-singular regions is based on the departure of the real matrix element from its soft and/or
collinear approximations.
For the plots in Fig.~\ref{fig:Z_vs_PWG} we adopt the
\pwgg{} option {\tt bornzerodamp=0}, enabling the exponentiation of the full NLO real matrix element. With this setting the \pwg{} tail gets accidentally close to the orange \rad{}+\matr{} curve for $ 50 \lesssim  p_t^{\mu\mu} \lesssim 150$ GeV, although not featuring any exact NNLO information contained in the latter, before reducing to the NLO result at larger $p_t^{\mu\mu}$. The matching systematics associated with the arbitrariness of the damping factor is not included in the \pwg{} uncertainty, which is obtained with a standard 7-point variation of the $\muR$ and $\muF$ scales. The relative smallness of the quoted \pwg{} band is then partially driven by missing information on resummation ($Q$) and matching ($p_{t0}$) uncertainties. Moreover, scale variations in \pwg{} have an effect only at the level of Les Houches events
\cite{Binoth:2010xt,Alioli:2013nda}, and are not entirely propagated in the showering phase. We stress that such a feature is rather common in NLO+PS computations, which typically do not include uncertainties stemming from the variations of $\muR$ and $\muF$ within the parton shower. A comparison of \pwg{} with the most accurate \rad{}+\matr{} prediction (pink vs orange) highlights the importance of including higher-order QCD and mixed QCD-EW effects. The shape modifications they induce with respect to the \pwg{} state of the art significantly exceed the quoted uncertainty band for the latter, which is foreseen to have an impact on precision DY phenomenology.

Turning to the positively-charged muon transverse momentum $p_t^{\mu^+}$ in the upper-right panel of Fig.~\ref{fig:Z_vs_PWG}, we observe that the region around the jacobian peak at $p_t^{\mu^+}\simeq m_Z/2$ is fairly sensitive to multiple soft and collinear radiation, hence to resummation effects. The remarkable level of compatibility between \rad{}+\matr{} and \pwg{} results with similar physical content (purple vs pink curves, i.e.~NLO+NLL$^\prime$ in QCD and EW couplings) reflects the agreement of the two results at small $p_t^{\mu\mu}$, already noticed in the upper-left panel of Fig.~\ref{fig:Z_vs_PWG}. The conclusions drawn earlier for the comparison with the best \rad{}+\matr{} result (pink vs orange) apply for $p_t^{\mu^+}$ as well, with shape distortions up to $\pm15\%$ in the displayed range, and a significant reduction of theoretical uncertainty after inclusion of higher-order corrections.

The di-muon transverse mass $m_t^{\mu\mu}$ shown in the lower panel of Fig.~\ref{fig:Z_vs_PWG} follows the same pattern, with a good agreement of central predictions from \pwg{} and \rad{}+\matr{} (pink vs purple) at NLO+NLL$^\prime$ accuracy. Shape distortions induced by higher-order contributions (pink vs orange) are milder than for $p_t^{\mu^+}$, and solely concern the region below the jacobian peak, almost reaching $-10\%$ in the displayed range. The reduction of theoretical uncertainty is again consistent, across the entire range, and more than a factor of 2 at small $m_t^{\mu\mu}$.

\subsection{Charged-current Drell Yan}
\label{sec:CCDY}

\begin{figure}[htbp]
\centering
\includegraphics[width=0.45\linewidth]{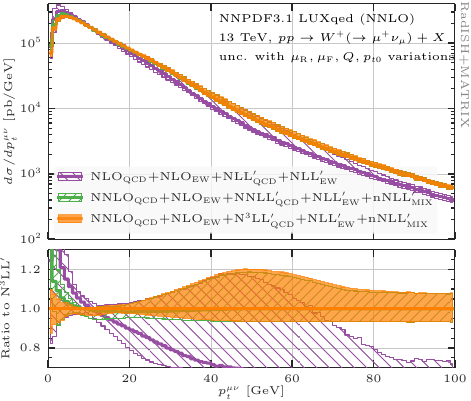}
\qquad
\includegraphics[width=0.45\linewidth]{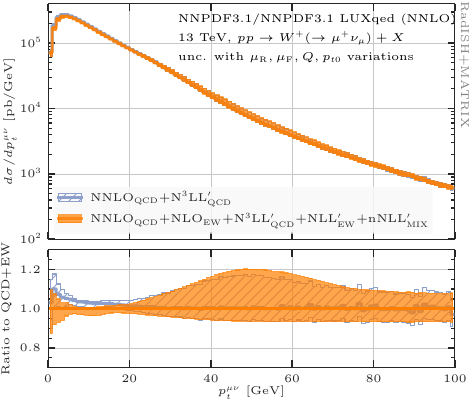}
\caption{Matched spectra for the muon-neutrino transverse momentum in charged-current DY. Left panel: perturbative progression including QCD and EW effects. Right panel: effect of EW corrections on top of the QCD baseline.}
\label{fig:WptB}
\end{figure}

We now turn to predictions for CCDY at the LHC, starting in Fig.~\ref{fig:WptB} with the muon-neutrino transverse momentum $p_t^{\mu\nu}$. Given the similarity of this observable with $p_t^{\mu\mu}$ in NCDY, the pattern displayed in Fig.~\ref{fig:WptB} is fairly similar qualitatively to that in Fig.~\ref{fig:ZptB}. From the left panel, we just remark a slightly larger residual uncertainty band with respect to NCDY, reaching the 25\% level in the matching region for our best prediction (orange) at $p_t^{\mu\nu}\simeq 50$ GeV. In the resummation region the uncertainty decreases to the few-\% level, showing clear perturbative convergence. In the right panel, the inclusion of EW corrections is again responsible for lowering the spectrum with respect to the QCD baseline below 20 GeV. In this case, the effect is quantitatively smaller than for NCDY, at the level of 5-10\% at most, compatibly with the presence of a single (as opposed to two) source of QED radiation in the CCDY Born final state.

\begin{figure}[htbp]
\centering
\includegraphics[width=0.45\linewidth]{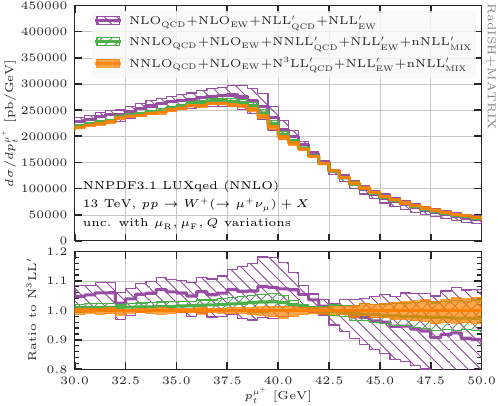}
\qquad
\includegraphics[width=0.45\linewidth]{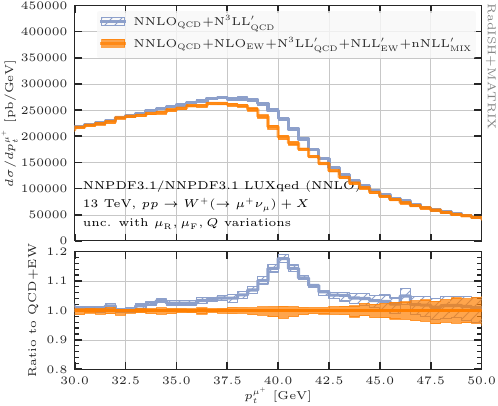}
\caption{Matched spectra for the muon transverse momentum in charged-current DY. Left panel: perturbative progression including QCD and EW effects. Right panel: effect of EW corrections on top of the QCD baseline.}
\label{fig:Wptlep}
\end{figure}

\begin{figure}[htbp]
\centering
\includegraphics[width=0.45\linewidth]{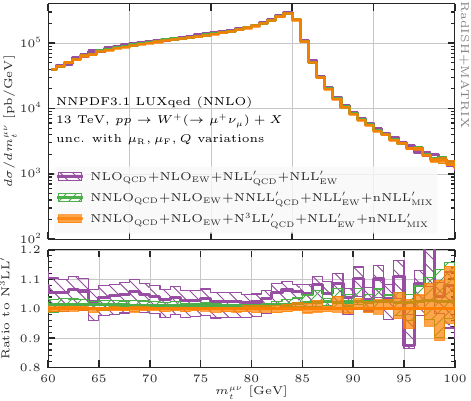}
\qquad
\includegraphics[width=0.45\linewidth]{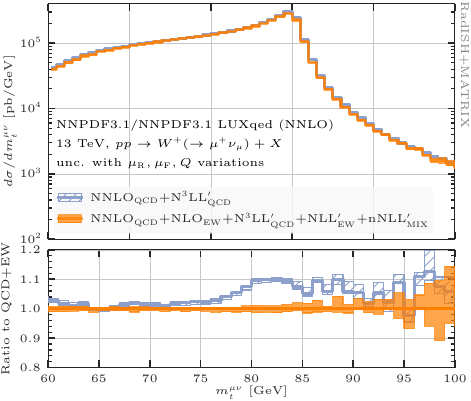}
\caption{Matched spectra for the muon-neutrino transverse mass in charged-current DY. Left panel: perturbative progression including QCD and EW effects. Right panel: effect of EW corrections on top of the QCD baseline.}
\label{fig:WmTr}
\end{figure}

Figs.~\ref{fig:Wptlep} and \ref{fig:WmTr} show predictions for the muon transverse momentum $p_t^{\mu^+}$ and for the muon-neutrino transverse mass $m_t^{\mu\nu}$. These distributions are central for the determination of fundamental SM parameters such as the $W$-boson mass, serving as inputs for template-fitting techniques \cite{Group:2012gb,ATLAS:2017rzl,LHCb:2021bjt,CDF:2022hxs,ATLAS:2023fsi}, or for the definition of new observables \cite{Rottoli:2023xdc,Torrielli:2023tiz} based on their perturbative prediction. By and large, the same comments expressed for the analogous NCDY observables apply in CCDY as well, with a remarkable perturbative stability displayed by all predictions including EW effects (left panels of Figs.~\ref{fig:Wptlep} and \ref{fig:WmTr}), and visible shape distortions induced by the latter on top of pure-QCD predictions (right panels of Figs.~\ref{fig:Wptlep} and \ref{fig:WmTr}). From the quantitative point of view, the effect of EW corrections is slightly smaller than for NCDY, consistently with what noticed for $p_t^{\mu\nu}$ in the right panel of Fig.~\ref{fig:WptB}. The trend is also very similar to what found in Fig.~24 of \cite{Alioli:2016fum}, both for $p_t^{\mu^+}$ and for $m_t^{\mu\nu}$.
\\

\begin{figure}[htbp]
\centering
\includegraphics[width=0.43\linewidth]{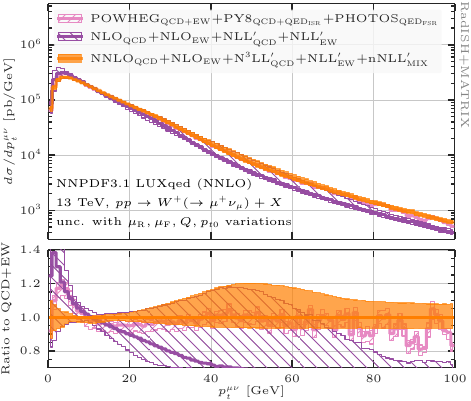}\qquad
\includegraphics[width=0.45\linewidth]{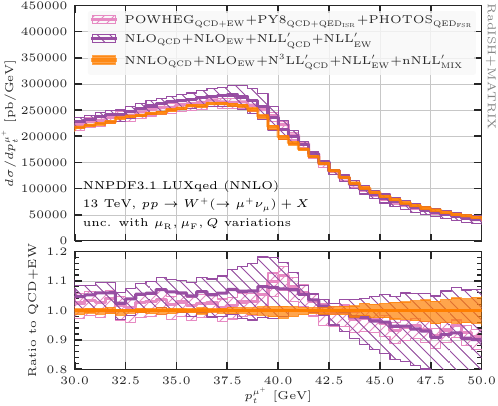}
\includegraphics[width=0.45\linewidth]{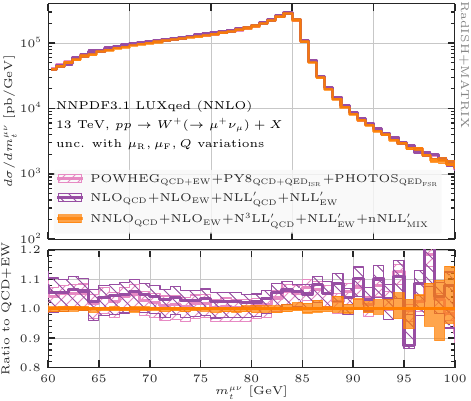}
\caption{Comparison of matched \rad{}+\matr{} spectra (purple and orange) against \pwg{} predictions (pink) for the muon-neutrino transverse momentum, the muon transverse momentum, and the muon-neutrino transverse mass in charged-current Drell Yan.}
\label{fig:W_vs_PWG}
\end{figure}

As for the comparison with \pwg{} in CCDY, in Fig.~\ref{fig:W_vs_PWG} we show predictions for the muon-neutrino transverse momentum, the muon transverse momentum, and the muon-neutrino transverse mass, with the same pattern used in Fig.~\ref{fig:Z_vs_PWG}. The features of the comparison are very similar, both qualitatively and quantitatively, to the ones already exposed in detail for NCDY, thus we refrain from further commenting on them. Given the high phenomenological relevance for these observables, we are confident that our new \rad{}+\matr{} predictions with highest accuracy (orange curves) will have an impact on the precise determination of the $W$-boson mass and the EW mixing angle at the LHC.

\subsection{Comparison between neutral- and charged-current Drell Yan}
\label{sec:CC_vs_NC}

We conclude the section of phenomenological results by showing the normalised ratio of the CCDY to NCDY di-lepton transverse momentum $p_t^{\ell\ell}$. This is a crucial control observable in the experimental strategy for $W$-boson mass extraction at the LHC \cite{ATLAS:2017rzl}. The differential spectra are normalised to the fiducial cross sections in the range $p_t^{\ell\ell} \in [0,30]$ GeV.

\begin{figure}[htbp]
\centering
\includegraphics[width=0.45\linewidth]{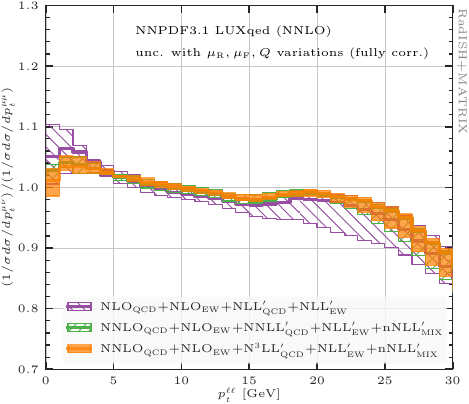}
\qquad
\includegraphics[width=0.45\linewidth]{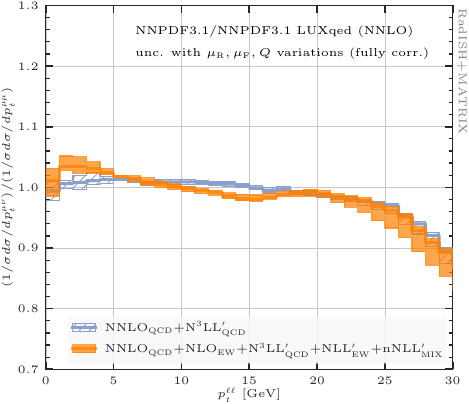}
\caption{Normalised ratio of charged- to neutral-current Drell Yan di-lepton transverse momentum. Variations of $\muR$, $\muF$, and $Q$ are correlated between the numerator and the denominator of the ratio. Left panel: perturbative progression including QCD and EW effects. Right panel: effect of EW corrections on top of the QCD baseline.}
\label{fig:ZWratio_corr}
\end{figure}

In Fig.~\ref{fig:ZWratio_corr}, using the same pattern as in the previous figures, we display \rad{}+\matr{} predictions for the ratio observable. We do not consider variations of the matching scheme, i.e. we set $Z(p_t^{\ell\ell})=1$ in \eq{eq:add_match2}.
Uncertainty bands are obtained with a fully correlated variation of the three perturbative scales $\muR$, $\muF$, and $Q$ in the numerator and in the denominator.
From the left panel of Fig.~\ref{fig:ZWratio_corr} we observe a robust perturbative progression in presence of EW effects, with higher-order corrections stably contained into uncertainty bands of lower orders. There is a significant uncertainty reduction (green vs purple) upon inclusion of NNLO+NNLL$^\prime$ QCD corrections and nNLL$^\prime_{\rm MIX}$ effects, while the further addition of N$^3$LL$^\prime$ QCD resummation (orange vs green) yields a more marginal reduction, essentially confirming the shape obtained at the previous QCD logarithmic accuracy. The shape itself is relatively non-trivial, as due to the interplay of EW corrections from initial- and final-state radiation with the fiducial cuts adopted.
The right panel of Fig.~\ref{fig:ZWratio_corr} shows the impact of EW corrections (orange) on top of the QCD baseline (light blue). The main distortion is observed at small $p_t^{\ell\ell}$, compatibly with what was noticed in the individual di-lepton transverse momentum spectra in Fig.~\ref{fig:ZptB} and Fig.~\ref{fig:WptB}. EW effects increase the slope of the ratio at $p_t^{\ell\ell}\lesssim 15$ GeV, reaching the level of $\pm 3\%$, and exceeding the QCD theoretical uncertainty band.
We note that the impact of EW corrections on the ratio observable is significantly more pronounced than the $\pm0.5\%$ observed in Fig.~6 of \cite{Autieri:2023xme}. Apart from differences in the setup and in the perturbative accuracy, the bulk of the discrepancy is due to fact that the analysis of \cite{Autieri:2023xme} is performed with undecayed $Z$ and $W$ gauge bosons, and inclusively over their phase space. We have checked that the ratio with EW effects indeed gets much closer to the pure QCD result upon removing the effect of QED radiation off final-state leptons. This highlights once more the importance of working with leptons at the fiducial level for precision DY phenomenology.

\begin{figure}[htbp]
\centering
\includegraphics[width=0.45\linewidth]{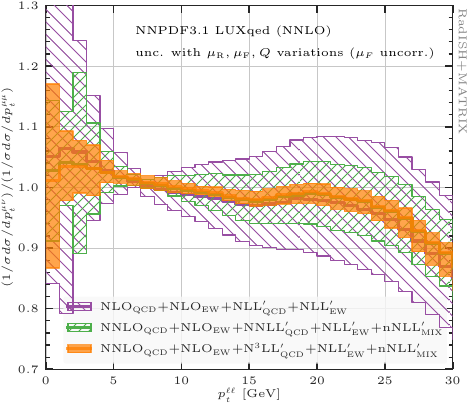}
\qquad
\includegraphics[width=0.45\linewidth]{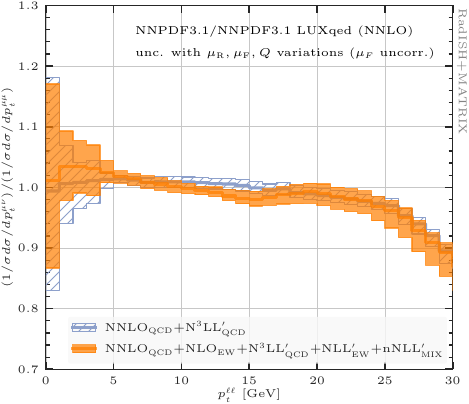}
\caption{Normalised ratio of charged- to neutral-current Drell Yan di-lepton transverse momentum. Variations of $\muR$, and $Q$ are correlated between the numerator and the denominator of the ratio, while variations of $\muF$ are only constrained by $1/2 \leq \muF^{\rm num}/\muF^{\rm den} \leq 2$. Left panel: perturbative progression including QCD and EW effects. Right panel: effect of EW corrections on top of the QCD baseline.}
\label{fig:ZWratio_mufuncorr}
\end{figure}

In Fig.~\ref{fig:ZWratio_mufuncorr} the CCDY to NCDY ratio is shown with a more conservative assumption on the correlation of scale variations. In particular, while the renormalisation and resummation scales are still varied in a fully correlated fashion, the factorisation scales for the numerator ($\muF^{\rm num}$) and for the denominator ($\muF^{\rm den}$) are varied independently, with the sole constraint $1/2 \leq \muF^{\rm num}/\muF^{\rm den} \leq 2$. This uncertainty prescription was already introduced in \cite{Gehrmann-DeRidder:2017mvr,Bizon:2019zgf}, and is physically motivated by considering that CCDY and NCDY probe different combinations of partonic channels, and of PDFs in turn, hence full $\muF$ correlation may not be clearly justified. Decorrelating $\muF$ variations causes a significant inflation in uncertainty bands, especially at small $p_t^{\ell\ell}$ and for predictions with lower formal accuracy, as seen comparing the left panels of Fig.~\ref{fig:ZWratio_mufuncorr} and of Fig.~\ref{fig:ZWratio_corr}.
As a result of this more conservative uncertainty estimate, predictions with and without EW effects in the right panel of Fig.~\ref{fig:ZWratio_mufuncorr} are now marginally compatible.

\begin{figure}[htbp]
\centering
\includegraphics[width=0.45\linewidth]{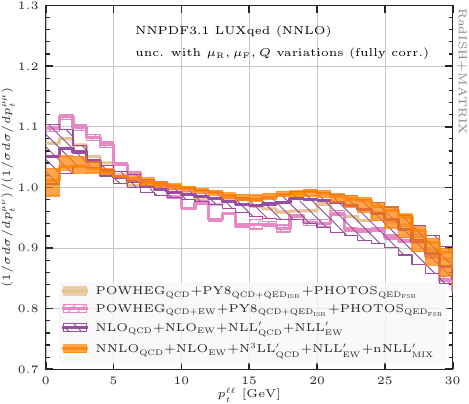}
\qquad
\includegraphics[width=0.45\linewidth]{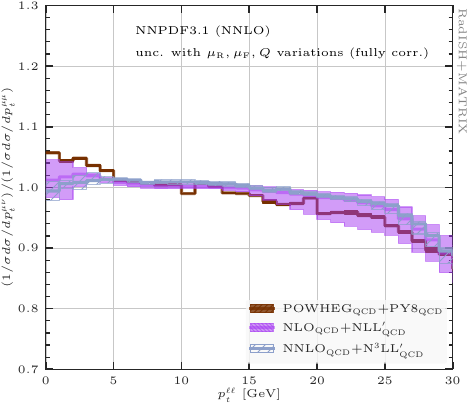}
\caption{Comparison between \rad{}+\matr{} and \pwgg{} for the normalised charged- to neutral-current Drell Yan di-lepton transverse momentum. Fully correlated variations of $\muR$, $\muF$, and $Q$ are considered. Left panel: predictions including QCD as well as EW effects. Right panel: predictions including solely QCD effects.}
\label{fig:ZWratio_vs_PWG}
\end{figure}

Finally, Fig.~\ref{fig:ZWratio_vs_PWG} reports the comparison of \rad{}+\matr{} and \pwgg{} predictions for the ratio observable, including QCD and EW contributions (left panel), or solely QCD effects (right panel).
Although the \pwg{} predictions for individual $p_t^{\ell \ell}$ distributions are in reasonable agreement with the NLO+NLL$^\prime$ \rad{}+\matr{} ones, the left panel of Fig.~\ref{fig:ZWratio_vs_PWG} reveals a moderate shape discrepancy in the ratio (purple vs pink), with \pwg{} being steeper in the displayed range. The discrepancy is not covered assuming fully correlated uncertainties, and only upon decorrelating uncertainties do the two predictions become compatible with each other, due to the large uncertainty of the purple band in Fig.~\ref{fig:ZWratio_mufuncorr}. As for the comparison of \pwg{} with our most accurate predictions (pink vs orange), scale decorrelation is not sufficient to cover the difference between the two curves across the whole range.
To further investigate this discrepancy, in the left panel of Fig.~\ref{fig:ZWratio_vs_PWG} we show in light brown a second \pwgg{} prediction, where QCD and QED showers are applied to a pure-QCD NLO sample obtained with the event generator of Ref.~\cite{Alioli:2008gx}. The behaviour in this case is in good agreement with the \rad{}+\matr{} NLO+NLL$^\prime$ prediction (light brown vs purple). The effect caused by NLO EW matrix elements in \pwg{} warrants further investigation, which however is beyond the scope of this article.

The \pwgg{} vs \rad{}+\matr{} pattern at the level of pure QCD, displayed in the right panel of Fig.~\ref{fig:ZWratio_vs_PWG}, is relatively similar to the light brown vs purple comparison in the left panel. The \pwgg{} prediction (dark brown) is slightly steeper, but still largely contained in the \rad{}+\matr{} NLO+NLL$^\prime$ correlated uncertainty band (lilac).
Although we did not perform a detailed study, we have checked that the \pwgg{} prediction for the ratio is relatively insensitive to the employed \py{} tune, which hints at a perturbative explanation for the remaining differences.
These shape features are qualitatively similar to the pattern observed in Ref.~\cite{ATLAS:2024nrd}, where a comparison for the ratio observable between \pwgg{} and \rad{} can be deduced. There, \pwgg{} is found to describe data slightly better than \rad{}, despite its substantially lower formal accuracy.
However, we note that in Ref.~\cite{ATLAS:2024nrd} the dominant effect of EW final-state radiation has been subtracted from the data, which moreover do not include any photon-induced contribution. The predictions presented in this article would allow for an accurate comparison including all sources of EW effects at the level of bare muons, as well as to establish more robustly the reliability of the EW subtraction procedure widely adopted in experimental analyses.

\section{Conclusion}
\label{sec:end}
In this article we have presented an extension of the \rad{} resummation
framework to include dominant classes of QED, virtual EW and mixed QCD-EW
effects in neutral- and charged-current Drell Yan lepton-pair production
featuring massive bare leptons.
Our predictions reach next-to-leading-logarithmic accuracy in the mixed QCD-EW coupling expansion, namely they correctly incorporate all contributions of order $\as^n \aw^m L^{n+m}$, in terms of the QCD and EW
coupling constants $\as$ and $\aw$, and of the large resummed logarithms $L$.
They also include all terms, beyond next-to-leading logarithms, necessary to
perform a consistent matching with fixed-order predictions at order ${\cal O}(\as \aw)$ relative to the Born level, a development that we leave for future work.

The predictions presented here extend in several directions the current state of the art \cite{Cieri:2018sfk,Autieri:2023xme} for the analytic resummation of mixed QCD-EW effects in Drell Yan.
First, our resummation features further subleading terms with respect to those considered in \cite{Cieri:2018sfk,Autieri:2023xme}, in particular all contributions with a ``(1,1)'' label in eqs.~(\ref{eq:finalR}) to (\ref{eq:PEW}), necessary for matching at order $\as \aw$. 
Second, the resummation formalism is not limited to the di-lepton transverse momentum, but can be applied with no modifications to all observables resummed by \rad{}, a notable example being $\phi_\eta^*$ in neutral-current Drell Yan. This can also open the door to the exploration of EW effects in different resummation environments still available in \rad{}, such as jet-vetos or double-differential resummations, or for other scattering processes.
Third, and most important, our predictions are fully differential in the phase space of the Drell Yan final-state leptons.
Working with the leptonic final state enables a consistent account of off-shell and interference effects, as well as the inclusion of non-resonant channels, such as the photon-initiated Born contribution in neutral-current Drell Yan. Moreover, it enables a physical description of final-state QED photon radiation, which in our results is included both in charge- and in neutral-current Drell Yan. Being fully differential allows us to obtain predictions for leptonic Drell Yan observables of phenomenological interest, as well as to apply fiducial selection cuts to match experimental analyses.
These features give our framework a level of flexibility
comparable to the one of dedicated EW Monte Carlo generators \cite{Golonka:2005pn,CarloniCalame:2003ux,CarloniCalame:2005vc,Barze:2012tt,Barze:2013fru}, with the advantage of retaining a higher formal accuracy in the all-order resummation. We expect our predictions to have a direct impact on high-precision Drell Yan phenomenology, especially for the determination of fundamental Standard Model parameters such as the $W$-boson mass and the EW mixing angle.

We have displayed the impact of EW effects on physical distributions in neutral-
and charged-current Drell Yan at the 13 TeV LHC. The perturbative
behaviour is robust for all predictions including EW effects. Scale uncertainties affecting our most accurate QCD-EW results are at the level of 2-5\% for inclusive observables such as the charged-lepton transverse momentum and the di-lepton transverse mass. The di-lepton transverse momentum has instead uncertainties that range from few-\% in the resummation region, to
15-20\% in the region where the transition to the fixed-order regime takes place.

We have been able to perform a meaningful comparison with EW Monte Carlo tools available in the literature. To this aim, we have used \pwg{} \cite{Barze:2012tt,Barze:2013fru}, which is the current state of the art for the resummation of the dominant QED effects in Drell Yan at leptonic level, and is used in Drell Yan experimental analyses. As for our NLO+NLL$^\prime$ QCD and EW predictions, the shape agreement with \pwg{} is overall good, however we have argued that our estimate of theoretical uncertainties is more robust. Higher-order QCD and EW effects included in our predictions result in shape distortions in the resummation-dominated kinematical regions, which may in turn
impact precision Drell Yan phenomenology.

In conclusion, accounting for EW effects on top of pure-QCD predictions causes modifications in the physical distributions that often exceed the quoted QCD theoretical uncertainty. This is for instance visible at small di-lepton transverse momentum, and around the jacobian peaks of the charged-lepton transverse momentum and of the transverse mass, with effects reaching 15-20\%.  This consideration highlights the importance of a complete description of Standard Model effects, not limited to QCD, for a successful precision-physics programme with Drell Yan observables at hadron colliders.

\section*{Acknowledgements}
We thank Massimiliano Grazzini, Pier Francesco Monni, Emanuele Re, and Alessandro Vicini for a careful reading of the manuscript. The work of LR has been supported by the SNSF under contract PZ00P2 201878. PT has been partially supported by the Italian Ministry of University and Research (MUR) through grant
PRIN 2022BCXSW9, and by Compagnia di San Paolo through grant
TORP\_S1921\_EX-POST\_21\_01. The work of LB is funded by the European Union (ERC, grant agreement No. 101044599, JANUS). Views and opinions expressed are however those of the authors only and do not necessarily reflect those of the European Union or the European Research Council Executive Agency. Neither the European Union nor the granting authority can be held responsible for them.

\appendix

\section{Formul\ae}
\label{sec:formulae}
In this section we report analytic elements relevant to the formul\ae{} used in the main text.

The QCD beta function reads
\beq
\frac{\rd\as(\mu)}{\rd\ln \mu^2}
\, = \,
\beta(\as,\aw)
\, \equiv \,
- \, \as
\left( \,
\beta_0 \, \as +
\beta_1 \, \as^2 +
\beta_{01} \, \aw +
\dots
\right)
\, ,
\eeq
whose first two coefficients (with $n_f$ active flavours, $C_A = N_c$, $C_F = \frac{N_c^2-1}{2 \, N_c}$, and $N_c = 3$) are
\beq
&&
\beta_0
\, = \,
\frac{11 \, C_A - 2 \, n_f}{12 \, \pi}
\, ,
\qquad
\beta_1
\, = \, 
\frac{17 \, C_A^{\,2} - 5 \, C_A \, n_f - 3 \, C_F \, n_f}{24 \, \pi^2}
\, .
\eeq

Next we report the functions entering the QCD Sudakov radiator up to NLL, with $\lambda = \as(\muR) \, \beta_0 \, L$, and $L=\ln\frac Q{k_{t1}}$. For Drell Yan, they read
\beq
\label{eq:g1g2}
g_{1}(\lambda)
& = &
\frac{A^{(1)}}{2\pi\beta_{0}}
\frac{2 \lambda +\ln (1-2 \lambda )}{2  \lambda }
\, ,
\nnb\\[4pt]
g_{2}(\lambda)
& = &
\frac{A^{(1)} \ln \frac{M^2}{Q^2}+B^{(1)}}{4\pi \beta_{0}}\ln (1-2 \lambda )
-
\frac{A^{(2)}}{8 \pi ^2 \beta_{0}^2}
\frac{2 \lambda +(1-2\lambda ) \ln (1-2 \lambda)}
{1-2\lambda}
\nnb\\
&&
-
\frac{A^{(1)} \beta_{1}}{8 \pi \beta_{0}^3}
\frac{\ln (1-2 \lambda )
\big[(2 \lambda -1) \ln (1-2 \lambda )-2\big]
-4\lambda}{1-2 \lambda}
\nnb\\
&&
-\frac{A^{(1)}}{4 \pi \beta_{0}}
\frac{2 \lambda
\big[1 - \ln (1-2 \lambda )\big]+\ln (1-2 \lambda )}{1-2\lambda} \ln
\frac{\mu_R^2}{Q^2}
\, ,
\eeq
with
\beq
&&
A^{(k)}
\, = \,
\sum_{\ell=1}^2
A_\ell^{(k)}
\, = \,
2 \, A_q^{(k)}
\, ,
\qquad
B^{(1)}
\, = \,
\sum_{\ell=1}^2
B_\ell^{(1)}
\, = \,
2 \, B_q^{(1)}
\, ,
\nnb\\
&&
A^{(1)}_q
\, = \,
2 \, C_F
\, ,
\qquad
A^{(2)}_q
\, = \,
2 \, C_F \, 
\bigg[
C_A
\Big(
\frac{67}{18}
-
\frac{\pi^2}{6}
\Big)
-\frac59 \, n_f
\bigg]
\, ,
\qquad
B^{(1)}_q
\, = \,
- \, 3 \, C_F
\, .
\eeq

The EW beta function reads
\beq
\frac{\rd\aw(\mu)}{\rd\ln \mu^2}
\, = \,
\beta'(\aw,\as)
\, \equiv \,
- \, \aw
\left( \,
\beta'_0 \, \aw +
\beta'_1 \, \aw^2 +
\beta'_{01} \as +
\dots
\right)
\, ,
\eeq
with
\beq
&&
\beta'_0
\, = \,
- \frac{N^{(2)}}{3 \, \pi}
\, ,
\qquad
\beta'_1
\, = \,
- 
\frac{N^{(4)}}{4 \, \pi^2}
\, ,
\qquad
N^{(k)}
\, = \,
N_c \, \sum_{q=1}^{n_f} e_{f_q}^k
+ \sum_{l=1}^{n_l} e_{f_l}^k
\, ,
\eeq
with $f_q=q$ and $f_l = 2l+9$ indicating quark and lepton flavours (following PDG conventions \cite{ParticleDataGroup:2022pth}), $e_{f_q}$ being quark electric charges ($+2/3$ for up-type, $-1/3$ for down type), $e_{f_l}=-1$ being lepton charges, and $n_l$ the number of lepton families considered.
\\

The building blocks $g_{1,2}'(\lambda')$ of the NLL EW Sudakov radiator are obtained from the corresponding QCD functions in \eq{eq:g1g2} with the formal replacements $\lambda\to\lambda' = \aw(\muR) \, \beta_0' \, L$, $\beta_k\to \beta_k'$, $A^{(k)}\to A^{\prime(k)}$, and $B^{(1)}\to \widetilde B^{\prime(1)}$. The relevant anomalous dimensions for Drell Yan are
\beq
\label{appA:anomalous_dimensions}
&&
A^{\prime(k)}
\, = \,
\sum_{\ell=1}^2
A_\ell^{\prime(k)}
\, ,
\qquad
\widetilde
B^{\prime(1)}
\, = \,
\sum_{\ell=1}^2
B_\ell^{\prime(1)}
+
D^{\prime(1)}(\Phi_B)
\, ,
\eeq
with~\cite{Cieri:2018sfk,Buonocore:2019puv,Cieri:2020ikq,Buonocore:2021rxx}
\beq
&&
A^{\prime(1)}_{\ell}
\, = \,
2 \, e_{f(\ell)}^2
\, ,
\qquad
A^{\prime(2)}_{\ell}
\, = \,
-
\frac{20}9 \, N^{(2)} \, e_{f(\ell)}^2
\, ,
\qquad
B^{\prime(1)}_{\ell}
\, = \,
- \, 3 \, e_{f(\ell)}^2
\, ,
\\
&&
D^{\prime(1)}(\Phi_B)
\, = \,
- \, 2 \, 
\bigg[
e_{f(3)} e_{f(4)}
\frac{1+\beta^2}\beta \, \ln\frac{1+\beta}{1-\beta} +
\sum_{\ell=1}^2
\sum_{k=3}^4
\Big(
\frac{e_{f(k)}^2}2 + e_{f(\ell)} e_{f(k)} \ln\frac{s_{\ell k}^2}{s_{12} \, m^2}
\Big)
\bigg]
\, ,
\nnb
\eeq
$f(j)$ the flavour of leg $j$, $s_{ij}=2p_i\cdot p_j$, $\beta=\sqrt{1-4 \, m^2/s_{12}}$, and $m$ the mass of the charged final-state lepton(s).

The $\mathcal{O}(\alpha)$ soft wide-angle contribution
reads \cite{Catani:2014qha,Buonocore:2020ren}
\beq F^{\prime(1)}(\Phi_B) =
\left(e_{f(3)}^2 + e_{f(4)}^2 \right)\ln \frac{m_{t}^2}{m^2} +
(e_{f(3)}+e_{f(4)})^2 {\rm Li}_2\left(\frac{-p_{t}^2}{m^2}\right) + e_{f(3)}
e_{f(4)} \frac{1}{v} L_{34} (v,y_{34}) \eeq with
$p_{t}\equiv p_{t,f(3)}=p_{t,f(3)}$ the common transverse momentum of the
leptons, $m_{t}=\sqrt{p_{t}^{2}+m^{2}}$ their transverse mass,
$y_{34} = y_{3}-y_{4}$ the rapidity difference between the two leptons and
\begin{align}
\label{L34}
L_{34}&=\ln\left( \frac{1+v}{1-v} \right)
\, \ln \left(\frac{m_t^2}{m^2}\right)
-2 \,{\rm Li}_2\left( \frac{2 v}{1+v}\right)
-\frac{1}{4}\ln^{2}\left(\frac{1+v}{1-v}\right)\notag \\
&+2\left[ \,{\rm Li}_2\left( 1 -
\sqrt{\frac{1-v}{1+v}}\, e^{\,y_{34}} \right)
+ \,{\rm Li}_2\left( 1 -
\sqrt{\frac{1-v}{1+v}}\,e^{-y_{34}} \right)
+ \frac{1}{2}\,y_{34}^2\right]
\end{align}
in terms of the relative velocity $v=\sqrt{1-4\, m^4/s_{34}^2}$.

The mixed QCD-EW radiator functions at lowest order are (see also \cite{Cieri:2018sfk})
\beq
g_{11}(\lambda,\lambda')
& = &
\frac{A^{(1)}\beta_{01}}{4\pi\beta_0'\beta_0^2}
\bigg[
\frac{\ln(1-2\lambda')}{1-2\lambda}
+\frac{\lambda'}{\lambda-\lambda'}\ln\frac{1-2\lambda'}{1-2\lambda}
+\ln(1-2\lambda') \ln\frac{(1-2\lambda)\lambda'}{\lambda'-\lambda}
\nnb\\
&&
\qquad\qquad
+ \, {\rm Li}_2\Big(\frac{\lambda(1-2\lambda')}{\lambda-\lambda'}\Big)
-{\rm Li}_2\Big(\frac{\lambda}{\lambda-\lambda'}\Big)
\bigg]
\, ,
\eeq
while $g'_{11}(\lambda,\lambda')$ is obtained from the above function with the replacements $\lambda \leftrightarrow \lambda'$, $\beta_j \leftrightarrow \beta_j'$, $A^{(1)} \to A^{\prime(1)}$.

Finally, the mixed hard-collinear anomalous dimension can be deduced abelianising the $B^{(2)}$ QCD coefficient~\cite{Buonocore:2020ren,Cieri:2020ikq}:
\beq
B^{(1,1)}
\, = \,
\sum_{\ell=1}^2
B^{(1,1)}_\ell
\, ,
\qquad
B^{(1,1)}_\ell
\, = \,
e_{f(\ell)}^2
\,
\frac83
\Big(
\pi^2 -
\frac34 -
12 \, \zeta_3
\Big)
\, .
\eeq

\bibliographystyle{JHEP}
\bibliography{mixed}

\end{document}